\documentclass[twocolumn,times]{aastex63}
\usepackage{amsmath}
\usepackage{newtxmath}
\usepackage{mleftright}
\mleftright

\newcommand{\appropto}{\mathrel{\vcenter{
  \offinterlineskip\halign{\hfil$##$\cr
    \propto\cr\noalign{\kern2pt}\sim\cr\noalign{\kern-2pt}}}}}
\shorttitle{Magnetic braking of accreting T Tauri stars}
\shortauthors{Ireland et al.}

\begin{document}

\title{Magnetic braking of accreting T Tauri stars: Effects of mass accretion rate, rotation, and dipolar field strength}

\author[0000-0002-8833-1204]{Lewis G. Ireland}
\affil{Department of Physics and Astronomy, University of Exeter, Stocker Road, Exeter, EX4 4QL, UK}

\author[0000-0003-0204-8190]{Claudio Zanni}
\affil{INAF - Osservatorio Astrofisico di Torino, Strada Osservatorio 20, {10025 Pino Torinese}, Italy}

\author[0000-0001-9590-2274]{Sean P. Matt}
\affil{Department of Physics and Astronomy, University of Exeter, Stocker Road, Exeter, EX4 4QL, UK}

\author[0000-0001-7788-3727]{George Pantolmos}
\affil{Univ. Grenoble Alpes, CNRS, IPAG, 38000 Grenoble, France}

\correspondingauthor{Lewis G. Ireland}
\email{L.G.Ireland@exeter.ac.uk}



\begin{abstract}


The rotational evolution of accreting pre-main-sequence stars is influenced by its magnetic interaction with its surrounding circumstellar disk. Using the PLUTO code, we perform 2.5D magnetohydrodynamic, axisymmetric, time-dependent simulations of star-disk interaction---with an initial dipolar magnetic field structure, and a viscous and resistive accretion disk---in order to model the three mechanisms that contribute to the net stellar torque: accretion flow, stellar wind, and magnetospheric ejections (periodic inflation and reconnection events). 
We investigate how changes in the stellar magnetic field strength, rotation rate, and mass accretion rate (changing the initial disk density) affect the net stellar torque. 
All simulations are in a net spin-up regime. 
We fit semi-analytic functions for the three stellar torque contributions, allowing for the prediction of the net stellar torque for our parameter regime, and the possibility of investigating spin-evolution using 1D stellar evolution codes.
The presence of an accretion disk appears to increase the efficiency of {stellar torques compared to isolated stars}, for cases with outflow rates much smaller than accretion rates, because the star-disk interaction opens more of the stellar magnetic flux compared to that from isolated stars. In our parameter regime, a stellar wind with a mass loss rate of $\approx 1 \%$ of the mass accretion rate is capable of extracting $\lesssim 50 \%$ of the accreting angular momentum. These simulations suggest that achieving spin-equilibrium in a representative T Tauri case within our parameter regime, e.g., BP Tau, would require a wind mass loss rate of $\approx 25\%$ of the mass accretion rate.

\end{abstract}




\section{Introduction}

Pre-main-sequence (PMS) solar-like stars, such as classical T Tauri stars (CTTS), exhibit rotational evolution that still lacks a comprehensive theoretical explanation. Many of these protostellar objects are observed to have a wide range of rotation rates, and a majority of these appear to be rotating at speeds much lower than their break-up velocity \citep[see, e.g.,][]{1993A&A...272..176B,2004AJ....127.1029R,2007prpl.conf..297H}. 
CTTS are observed to still actively accrete material from a circumstellar accretion disk, exhibiting mass accretion rates of $\sim 10^{-10}-10^{-8}$ $M_\odot$ yr$^{-1}$  \citep{1994AJ....108.1056E,1998ApJ...492..323G,1998ApJ...495..385H,2014ApJ...790...47I,2017A&A...600A..20A}. Combined with the fact that these objects are still undergoing gravitational contraction, one would expect these protostars to be spinning up; however, observations have suggested solar-like stars in the PMS phase appear to have a fairly constant distribution of spin rates for a few million years \citep{1993AJ....106..372E,1993A&A...272..176B,2004AJ....127.1029R,2009IAUS..258..363I,2019A&A...632A...6G}. Therefore, to explain the rotational evolution we observe, we require an additional mechanism that efficiently removes angular momentum from the star during their PMS phase.

CTTS are magnetically active objects, exhibiting $\sim$ kG strength multipolar magnetic fields \citep[see, e.g.,][]{2007ApJ...664..975J,10.1111/j.1365-2966.2008.13687.x,2014MNRAS.437.3202J,2019MNRAS.483L...1D,2020MNRAS.491.5660D}, and it has been theorized that magnetic interaction between a star and its disk allows for the removal of angular momentum, allowing accreting stars to achieve slow rotation rates \citep[see, e.g.,][]{1990RvMA....3..234C,1991ApJ...370L..39K}. 
The classical \citet{1979ApJ...234..296G} model proposes that a spin-down torque is exerted onto the stellar surface from magnetic field lines connecting the star and the accretion disk beyond the corotation radius, where the disk is rotating more slowly than the stellar rotation rate. However, this assumption that the magnetic field is predominantly closed and connected to a large proportion of the disk is not consistent with our modern picture. It has been shown both analytically and via magnetohydrodynamical (MHD) simulations that the ``twisting" of magnetic field lines as a result of differential rotation between the star and the disk inflate and open the magnetosphere at mid-latitudes \citep[see][and references therein]{2005MNRAS.356..167M}. Combined with the dilution of the poloidal field intensity \citep{2000MNRAS.317..273A,2009A&A...508.1117Z}, the efficiency of the spin-down torque associated with the extended magnetosphere in the \citet{1979ApJ...234..296G} picture is vastly reduced.

Other proposed solutions consider the presence of outflows, allowing angular momentum to be removed from the system. Magnetized stellar winds extract angular momentum from the stellar surface via open magnetic field lines, providing a spin-down torque contribution \citep{2005ApJ...632L.135M}. \citet{2008ApJ...681..391M} show that for spin-equilibrium, where the spin-up accretion torque balances with the spin-down stellar wind torque, the stellar wind mass loss rate is likely $\sim 10 \%$ of the mass accretion rate. However, this is unlikely achievable if the outflows are predominantly powered by the star's rotational, thermal, and magnetic energies alone; therefore an additional energy source would be required that is not of thermal origin \citep{2007IAUS..243..299M}. 
The idea of accretion-powered stellar winds was proposed by \citet{2005ApJ...632L.135M}.  However, possible mechanisms behind the {power transfer have only} been explored in a few papers \citep{2008ApJ...689..316C,2009ApJ...706..824C}, and the energy available has strict limits \citep{2008ApJ...681..391M,2011ApJ...727L..22Z}.

The opening and reconnecting of closed magnetospheric field lines, coined magnetospheric ejections (MEs), is another ejection mechanism, and is a result of the build-up of toroidal magnetic field pressure due to the star-disk differential rotation \citep{refId0}. Performing 2.5D axisymmetric star-disk interaction (SDI) simulations, \citet{refId0} found that similar to a magnetic slingshot, MEs are powered by the star-disk differential rotation, and can extract angular momentum from both the disk and the star. 
In fact, they find it possible for a weak stellar wind torque, with a mass outflow rate $\sim 1\%$ of the mass accretion rate, to balance the spin-up accretion torque.
{In general, MEs take a net amount of angular momentum (from the disk) away from the system, reducing the accretion torque. As they are magnetically connected (intermittently) to both the star and the disk, a fraction of the extracted angular momentum can be given to the star. This happens when the disk is truncated close enough to the star such that the disk's orbital motion is faster than the stellar rotation. In that case, the magnetic field lines anchored in the disk rotate faster than the star, making them twist forward, with respect to the stellar rotation, at the stellar surface (but twisted backward at the disk surface, with respect to the disk's rotation). This imparts a spin-up torque on the star (and spin-down on the disk). In this case, the MEs can also add mass back onto the star; most of the stellar ME mass flux comes from the disk and can be sensibly larger than 1\% of the mass accretion rate \citep[see][]{refId0}. Alternatively, when the disk is truncated far enough from the star {such that the inner disk's angular rotation} is slower than the star, the magnetic field lines twist backwards, with respect to the stellar rotation, at the stellar surface, imparting a spin-down torque on the star. Therefore, the ME stellar torque contribution is primarily a function of whether the MEs anchor to the disk in a place where the disk rotates faster or slower than the star.}
Thus, MEs may be important in the spin evolution of PMS stars.

{The relevance of outflows powered by magnetospheric accretion, either of stellar or disk origin, on the spin evolution of CTTS {seems} to be supported by the fact that, on average, non-accreting or faintly accreting weak-lined T Tauri stars (WTTS) rotate faster than CTTS \citep[see, e.g.,][]{refId0_venuti}. While the presence of strong magnetospheric winds in CTTS is supported by the observation of P-Cygni line profiles {\citep[see, e.g.,][]{Edwards_2006}} that can be reproduced either by stellar wind or inner disk-wind models \citep{10.1111/j.1365-2966.2011.19216.x}, there are no reliable detections of winds in WTTS.}

We attempt to characterize the angular momentum problem over a parameter regime that may represent stars over a wide range of ages, and to further explore the ME mechanism and their interaction with stellar winds, using SDI simulations to extract torque prescriptions that can be used in secular spin-evolution models. 
Over the last decade, stellar wind torque prescriptions have been proposed \citep[see, e.g.,][]{2008ApJ...678.1109M,2009MNRAS.392.1022U,2012ApJ...754L..26M,2015ApJ...798..116R,2017ApJ...845...46F,2018ApJ...854...78F,2017ApJ...849...83P} using isolated stellar wind (ISW) MHD simulations. 
The first spin-evolution calculation to use stellar torque prescriptions derived from SDI simulations was \citet{2019A&A...632A...6G}, using simulations from \citet{refId0}, which they adopted in their spin evolution models. 
They found that a strong stellar wind with a mass outflow rate corresponding to $\approx 10 \%$ of the accretion rate, or a disk with a truncation radius close to its corotation radius, was required for an efficient spin-down torque configuration. They propose an accretion torque formulation that accounts for a fraction of the analytical value due to the extraction of angular momentum from the disk via MEs, and a stellar ME torque prescription that depends on the torque exerted by magnetic field lines coupled to the MEs, as a result of the differential rotation between the star and the accretion disk. However, their prescription was only based on four simulations, and employs coefficients that appear to be guided by these simulations, rather than explicitly fitted. Furthermore, they use a stellar wind torque prescription calibrated using ISW simulations only \citep{2012ApJ...754L..26M}. 

In this paper, we make further improvements to this torque prescription, to better represent a wide parameter regime of SDI simulations and capture the interplay between stellar winds and SDI, in a dynamically self-consistent way. {We introduce a new stellar wind torque formulation that encapsulates the effects of SDI, illustrating that the stellar wind (open) magnetic flux is strongly dependent on disk properties, predominantly the truncation radius. We propose an accretion torque formulation that depends on how the accretion disk deviates from Keplerian rotation for each simulation. We also propose an improved stellar ME torque prescription that again considers how strongly sub-Keplerian the disk is, but also accounts for the perturbation of the magnetosphere from simple dipolar scaling due to the existence of an accretion disk.}

We perform 38 2.5D axisymmetric SDI simulations, changing the initial disk density, the rotation rate, and the magnetic field strength, in order to capture how the net stellar torque, due to the stellar wind, accretion, and MEs, depends on both stellar and accretion disk properties. 
Our simulations also adopt an equation of state with a variable polytropic index, allowing us to simultaneously treat the cooler disk gas with a more physically meaningful $\gamma = 5/3$, whilst also allowing for a Parker stellar wind with $\gamma = 1.05$. 
{We therefore model the stellar winds as thermally-driven outflows, {by setting suitable density and temperature profiles.} Despite its simplicity, this approach has two main caveats.  Firstly, it has been shown that in order to be compatible with the emission properties and the accretion power available in CTTS, a thermally-driven wind must possess a very low mass-loss rate \citep{2007IAUS..243..299M}. From this point of view, our driving mechanism is most likely unrealistic, but since our goal is to evaluate the extraction of angular momentum via the stellar wind, the torque does not much depend on the precise mechanism that drives the wind. Secondly, this approach does not allow us to self-consistently model the energetic coupling between the wind and the accretion flow, which is usually invoked to produce wind mass loss rates that are a substantial fraction of the accretion rate \citep{2005ApJ...632L.135M,2008ApJ...689..316C,2009ApJ...706..824C,2011ApJ...727L..22Z}. In our simulations, the mass loss rate and wind power can be set independently of the accretion luminosity by adjusting the wind boundary conditions. Our setup results in wind mass loss rates typically of the order of 1\% of the accretion rate, which is consistent with the mass flux predicted by models of stellar winds powered by the dissipation of MHD turbulence and waves excited by the inhomogeneous impact of the accretion funnels \citep{2008ApJ...689..316C,2009ApJ...706..824C}.} 

In Section~\ref{sec:setup}, we introduce the numerical simulation set up, the initial and boundary conditions, normalizations, and the simulation parameter study. 
In Section~\ref{sec:sims_chapter}, we list our SDI simulations, present a qualitative description of their behavior and how it depends on the parameters in our study, and introduce the global quantities used in our investigation, such as the mass flow rate, torque, and unsigned magnetic flux. 
In Section~\ref{sec:torque_formulation}, we introduce and fit semi-analytic formulations for the truncation radius and the torques from accretion, MEs, and stellar wind, which allows for the net stellar torque to be calculated. 
In Section~\ref{sec:dis_conc}, we discuss our findings and summarize our conclusions.
 
\section{Numerical Method}\label{sec:setup}

\subsection{Numerical setup}

The simulations in this paper are solved numerically using the following MHD equations, including resistive and viscous effects:

\begin{align}\label{eq:MHD}
\begin{gathered}
\frac{\partial \rho}{\partial t} + \boldsymbol{\nabla} \cdot (\rho \boldsymbol{v}) = 0, \\
\frac{\partial \rho \boldsymbol{v}}{\partial t} + \boldsymbol{\nabla} \cdot \left[\rho \boldsymbol{v} \boldsymbol{v} + \left(p + \frac{\boldsymbol{B} \cdot \boldsymbol{B}}{8 \pi}\right) \boldsymbol{I} - \frac{\boldsymbol{B}\boldsymbol{B}}{4\pi} - \boldsymbol{\tau} \right] = \rho \boldsymbol{g}, \\
\frac{\partial E}{\partial t} + \boldsymbol{\nabla} \cdot \left[\left(E + p + \frac{\boldsymbol{B} \cdot \boldsymbol{B}}{8 \pi}\right) \boldsymbol{v} - \frac{(\boldsymbol{v} \cdot \boldsymbol{B}) \boldsymbol{B}}{4 \pi} \right] \\
= \rho \boldsymbol{g} \cdot \boldsymbol{v} + \boldsymbol{\nabla \tau} \cdot \boldsymbol{v} - \frac{\boldsymbol{B}}{4\pi} \cdot (\boldsymbol{\nabla} \times \eta_\text{m} \boldsymbol{J}), \\
\frac{\partial \boldsymbol{B}}{\partial t} + \boldsymbol{\nabla} \times (\boldsymbol{B} \times \boldsymbol{v} + \eta_\text{m} \boldsymbol{J}) = 0.
\end{gathered}
\end{align}
The expressions in Equation~(\ref{eq:MHD}) represent the mass-continuity, momentum, energy, and magnetic induction equation, respectively. $\rho$ is the mass density, $\boldsymbol{v}$ is the plasma flow velocity, $p$ is the plasma thermal pressure, $\boldsymbol{B}$ is the magnetic field, $\boldsymbol{I}$ is the identity matrix, $\boldsymbol{g} = - (GM_\star/R^2) \boldsymbol{\hat{R}}$ is the gravitational acceleration ($G$ is the gravitational constant, $M_\star$ is the stellar mass, and $R$ is the spherical radius), $\boldsymbol{J} = \boldsymbol{\nabla} \times \boldsymbol{B} / 4\pi$ is the electric current, and $\eta_\text{m}$ is the magnetic resistivity. 
The magnetic diffusivity is defined as $\nu_\text{m} = \eta_\text{m} / 4\pi$. The total energy density $E$ is defined as

\begin{equation}\label{eq:energy}
	E = \rho u + \rho \frac{\boldsymbol{v} \cdot \boldsymbol{v}}{2} + \frac{\boldsymbol{B} \cdot \boldsymbol{B}}{8\pi},
\end{equation}
where $u(T)$ is specific internal energy as a function of temperature $T$, determined via our new adopted caloric equation of state for a calorically imperfect gas, where the specific heats and adiabatic index $\gamma$ are temperature-dependent; this is described in detail in the appendix of {\citet{2020arXiv200900940P}}. The hot stellar wind plasma will behave near-isothermally ($\gamma = 1.05$), and the cold accretion disk will behave adiabatically ($\gamma=5/3$). The viscous stress tensor $\boldsymbol{\tau}$ is defined as

\begin{equation}\label{eq:stress}
	\boldsymbol{\tau} = \eta_\text{v} \left[(\boldsymbol{\nabla}\boldsymbol{v}) + (\boldsymbol{\nabla}\boldsymbol{v})^{T} - \frac{2}{3}(\boldsymbol{\nabla} \cdot \boldsymbol{v}) \boldsymbol{I}\right],
\end{equation}
where $\eta_\text{v}$ is the dynamic viscosity. The kinematic viscosity is defined as $\nu_\text{v} = \eta_\text{v} / \rho$. 
The viscous and magnetic diffusive terms on the right hand side of the total energy equation (in the system of Equations~(\ref{eq:MHD})) correspond to the work of the viscous forces and diffusion of the magnetic energy. Dissipative viscous and Ohmic heating terms are omitted from the total energy equation to avoid potential runaway heating of the accretion disk, as a result of the lack of cooling radiative effects in our equation \citep{2009A&A...508.1117Z}.

{We also solve the following scalar equations:}

\begin{align}\label{eq:MHD_entropy}
\begin{gathered}
{\frac{\partial \rho s}{\partial t} + \boldsymbol{\nabla} \cdot (\rho s\boldsymbol{v}) = 0,} \\
{\frac{\partial \rho Tr}{\partial t} + \boldsymbol{\nabla} \cdot (\rho Tr\boldsymbol{v}) = 0,} \\
\end{gathered}
\end{align}
{where $s$ is the specific entropy (defined using the new equation of state), and $Tr$ is a passive tracer. The specific entropy can be used to monitor dissipation and heating from numerical integration of the total energy equation. $Tr$ is used to distinguish between disk material and the stellar wind plasma.}

{To numerically solve Equations~(\ref{eq:MHD})--(\ref{eq:MHD_entropy}), we employ a second-order Godunov method from the PLUTO code\footnote{PLUTO is freely available to download at \url{http://plutocode.ph.unito.it}} \citep{0067-0049-170-1-228}. The MHD equations are solved in the rotating frame of the star. We achieve spatial reconstruction of primitive variables using a combination of linear and parabolic interpolation. To compute intercell fluxes, we use the approximate HLLD Riemann solver \citep{2010ITPS...38.2236M}, computing only the deviation from the initial magnetic field. To advance the MHD equations in time, we employ a second order Runge-Kutta scheme. We use the hyperbolic divergence cleaning method \citep{2002JCoPh.175..645D} to control the divergence free constraint $\boldsymbol{\nabla} \cdot \boldsymbol{B} = 0$. A second-order finite difference approximation, integrated in time explicitly, is adopted to compute viscous and resistive terms.}

The simulations in this paper are 2.5D (2D domain with 3D vector components), and uses a ($R,\theta$) spherical coordinate system. We assume axisymmetry about the stellar rotation axis. Hereafter, we define the spherical radius using $R$ and the cylindrical radius using $r = R \sin{\theta}$. We use a logarithmic grid in the $R$ direction, covering $[1,50.756521]R_\star$ across 320 grid cells, where $R_\star$ is the stellar radius, which ensures that the cell sizes satisfy $\Delta R \sim R \Delta \theta$. We use a uniform grid in the $\theta$ direction, covering $[0,\pi]$ across 256 grid cells.

\subsection{Initial conditions}\label{sec:initial_conditions}

{The computational domain is initialized with a stellar corona, a viscous accretion disk, and a magnetic field configuration.}

{The stellar corona is initialized using density and pressure profiles from a 1D spherically symmetric, isentropic, transonic Parker wind solution. This solution is defined by its base density $\rho_\star$ and sound speed at the stellar surface $c_{\text{s},\star}$. The poloidal speed is set to zero.}

The Keplerian accretion disk is initialized adopting an $\alpha$ parameterization \citet{1973A&A....24..337S} for the viscosity. 
Neglecting inertial terms, the thermal disk pressure $p_\text{d}$ and density $\rho_\text{d}$ are determined via vertical hydrostatic equilibrium, and the disk toroidal velocity $v_{\phi,\text{d}}$ is derived via radial equilibrium. Assuming the thermal disk pressure and density are related via the polytropic relation $p_\text{d} \propto \rho_\text{d}^\gamma$, where $\gamma=5/3$, we write

\begin{align}\label{eq:rho_disk_initial}
\begin{gathered}
\rho_\text{d} = \rho_{\text{d,}\star} \left\{\frac{2}{5\epsilon^2}\left[\frac{R_\star}{R} - \left(1 - \frac{5 \epsilon^2}{2}\right) \frac{R_\star}{r}\right]\right\}^{3/2},\\
p_\text{d} = \rho_{\text{d,}\star} \epsilon^2 v_{\text{K},\star}^2 \left(\frac{\rho_\text{d}}{\rho_{\text{d,}\star}}\right)^{5/3},\\
v_{\phi,\text{d}} = \left(1 - \frac{5 \epsilon^2}{2}\right)^{1/2} \left(\frac{GM_\star}{r}\right)^{1/2}, \\
\end{gathered}
\end{align}
{where $\epsilon=c_\text{s,d}/v_\text{K}|_{\theta=\pi/2}$ is the disk aspect ratio evaluated at the disk midplane, $c_\text{s,d}=(p_\text{d}/\rho_\text{d})^{1/2}$ is the disk isothermal sound speed}, $v_\text{K} = (GM_\star/r)^{1/2}$ is the Keplerian velocity, and $\rho_{\text{d,}\star}$ and $v_{\text{K},\star}$ are the disk density and Keplerian velocity, respectively, evaluated on the midplane at $R_\star$. The kinematic viscosity of the disk, using an $\alpha$ parameterization \citep{1973A&A....24..337S}, is expressed as

\begin{align}\label{eq:visc_disk_initial}
{\nu_\text{v} = \frac{2}{3} \alpha_\text{v} \frac{c_\text{s,d}^2}{\Omega_\text{K}},}
\end{align}
{where $\Omega_\text{K} = (GM_\star/r^3)^{1/2}$ is the Keplerian angular velocity}, and the magnetic diffusivity of the disk is expressed as a function of the kinematic viscosity:

\begin{equation}\label{eq:diffusivity}
\nu_\text{m} = \frac{3}{2} \alpha_\text{m} \frac{\nu_\text{v}}{\alpha_\text{v}},
\end{equation}
where $\alpha_\text{m}$ is the anomalous diffusivity coefficient. Thus, the magnetic Prandtl number $\text{Pr}_\text{m} = \nu_\text{v}/\nu_\text{m} \equiv 2\alpha_\text{v}/3\alpha_\text{m}$. The radial accretion velocity is expressed as

\begin{align}\label{eq:r_rot_disk_initial}
{v_{R,\text{d}} = - \alpha_\text{v} \frac{c_\text{s,d}^2}{v_\text{K}} \sin{\theta}.}
\end{align}
{In this expression, we neglect the $\tau_{\theta\phi}$ component of the stress tensor in Equation~(\ref{eq:stress}), which avoids the likely unphysical backflow along the disk midplane, usually associated with 3D models of $\alpha$ accretion disks \citep[see, e.g.,][]{refId0_2002,refId0}. 
Equation~(\ref{eq:r_rot_disk_initial}) demonstrates that the inertial term due to radial accretion flow in the radial momentum equation is of order $\mathcal{O}(\alpha_\text{v}\epsilon^4)$, whereas the thermal pressure gradient is of order $\mathcal{O}(\epsilon^2)$; therefore, the inertial term can be neglected when deriving the disk equilibrium solution (Equation~(\ref{eq:rho_disk_initial})).
The passive tracer $Tr$ is initialized to one inside the disk and zero in the stellar corona.}

{Even though we use an $\alpha$ model to compute our initial disk structure, the thermal and density profiles of our initial disk are different from a Shakura \& Sunyaev $\alpha$ disk or the illuminated $\alpha$ disk by \citet{D_Alessio_1998}, since we made a simplifying adiabatic assumption to compute the disk thermal structure. Our disk is intended to be only a dynamical model that provides a source of mass and angular momentum accreting at a suitable rate.}

The magnetospheric field is initially a purely dipolar configuration, which is aligned to the rotation axis. 
The $R$ and $\theta$ magnetic field components are

\begin{equation}\label{eq:Br}
B_R(R,\theta) = B_\star \left(\frac{R_\star}{R}\right)^3 \cos{\theta}
\end{equation}
and

\begin{equation}\label{eq:Btheta}
B_\theta(R,\theta) = \frac{1}{2} B_\star \left(\frac{R_\star}{R}\right)^3 \sin{\theta},
\end{equation}
respectively, where $B_\star$ is the stellar surface polar magnetic field strength. 

{The disk is initially truncated, such that it is excluded from the region where the coronal magnetic stresses exceed the disk thermal pressure. 
We determine an initial truncation radius $R_\text{t,i}$, which corresponds to the radius at which the following implicit equation in the $R$ direction is true:}

\begin{equation}\label{eq:M_s_R_t_guess}
{m_\text{s} = \left\lvert \frac{B_\phi^+ B_{d,\theta=\pi/2}}{2\pi p_{d,\theta=\pi/2}} \right\rvert,}
\end{equation}
{where $m_\text{s}$ is the sonic Mach number of the accretion flow due to the large-scale magnetic field torque, $B_\phi^+$ is the theoretical toroidal field at the disk surface due to star-disk differential rotation, estimated as}

\begin{equation}\label{eq:B_phi_guess}
{B_\phi^+ = \frac{B_{d,\theta=\pi/2}}{\alpha_\text{m} \epsilon} \left[\left(\frac{R}{R_\text{co}}\right)^{3/2} - 1\right],}
\end{equation}
{where $R_\text{co} = (GM_\star/\Omega_\star^2)^{1/3}$ is the Keplerian corotation radius where the disk's Keplerian rotation equals the stellar angular velocity $\Omega_\star$, and $B_{d,\theta=\pi/2} = B_\star (R/R_\star)^{-3}$ and $p_{d,\theta=\pi/2}$ are the initial magnetic field strength and the disk thermal pressure both at the disk midplane, respectively. We set $m_\text{s}=1.5$, resulting in transonic accretion at the inner disk radius; a typical condition for accretion funnels to form \citep[see, e.g.,][]{2008A&A...478..155B}. The upper limit is set as $R_\text{t,i} < 0.8 R_\text{co}$. Once $R_\text{t,i}$ is determined, we set the disk pressure and density equal to zero (but initialize with the stellar corona solution) inside that radius.

Due to star-disk interaction, the initial disk structure is strongly modified; for $R < R_\text{t,i}$, we use the local sound speed to define the kinematic viscosity (Equation~(\ref{eq:visc_disk_initial})) and magnetic diffusivity (Equation~(\ref{eq:diffusivity})), which is smoothly matched to the initial disk sound speed for $R > R_\text{t,i}$, where disk structure is largely unaffected. 
We multiply the viscosity and diffusivity by the passive scalar $Tr$, so that only the accretion disk is viscous and resistive. 
For $B^2/8\pi p > 1$, the viscosity and diffusivity are set to zero, to ensure ideal MHD conditions in regions of strong magnetization, such as the accretion funnels and stellar winds. Magnetic surfaces that initially thread the disk are imposed to rotate at Keplerian angular velocity calculated at the disk anchoring radius, while those within $R_\text{t,i}$ are imposed to corotate with the star.}

\subsection{Boundary Conditions}\label{sec:boundary_conditions}

{The computational domain is enclosed by a boundary and augmented with ghost cells to allow boundary conditions to be implemented.} {An ``axisymmetric" inner and outer boundary for the $\theta$ coordinate is employed, which symmetrizes variables and flips the signs of both normal and $\phi$ components of vector fields across the boundary. At the inner boundary for the $R$ coordinate, there are two conditions: for {subsonic flow into the computational domain (out from the stellar boundary, i.e., stellar wind)}, we impose fixed density and thermal pressure profiles from the Parker wind solution in the ghost {cells}; for {supersonic flow from the computational domain (into the stellar boundary, i.e., accretion columns)},
density and thermal pressure are free to adjust to the values determined by the accretion funnel. In the latter, a power-law extrapolation along magnetic field lines is employed for the density and the thermal pressure is set assuming constant entropy along magnetic surfaces. Intermediate regions, i.e., subsonic outflow or a hydrostatic corona, use the sonic Mach number calculated in the first row of cells in the domain to linearly interpolate between the two boundary conditions. The $R$ magnetic field component is kept fixed to conserve the total stellar flux, whereas the $\theta$ component is free to change using a linear extrapolation. We also impose the poloidal velocity $v_\text{p}$ to be parallel to the poloidal magnetic field, along with the continuity of the axisymmetric MHD invariant $k = 4\pi \rho v_\text{p}/B_\text{p}$ along field lines, to ensure smooth inflow of stellar wind and supersonic infall via accretion funnels onto the stellar surface, without any shock generation.} Following \citet{2009A&A...508.1117Z}, we model the stellar surface as a perfect conductor that rotates at an angular velocity $\Omega_\star$, and impose a boundary condition on $B_\phi$ that forces magnetic field lines to rotate at the stellar rotation rate. The radial derivative of $B_\phi$, which is used to do the boundary extrapolation, is derived by replacing the local acceleration in the angular momentum equation with the following:

\begin{equation}\label{eq:B_phi_deriv}
\rho \frac{\partial r\Omega}{\partial t} = \rho \frac{r\Omega_\star + v_\text{p} B_\phi / B_\text{p} - r \Omega}{\Delta t},
\end{equation}
where $\Delta t$ is the Alfvén crossing time of one grid cell close to the stellar surface. The stellar rotation at the inner boundary is set as

\begin{equation}\label{eq:inner_v_phi}
v_\phi = r \Omega + v_\text{p} \frac{B_\phi}{B_\text{p}}.
\end{equation}

We impose a power-law extrapolation for density and thermal pressure, and a linear extrapolation for all other variables, at the outer boundary for the $R$ coordinate. For the toroidal magnetic field along open field lines anchored to the stellar surface, we impose a boundary condition similar to the approach adopted at the inner boundary, in order to avoid artificial torques being exerted on the star due to sub-Alfvénic matter crossing the outer boundary.

\subsection{Units and normalization}\label{sec:units_norm}

Simulations in this paper are performed in dimensionless units. Here, we list normalization factors required to express our results in physical units. Length is expressed in units of the stellar radius $R_\star$, density is expressed in units of the initial coronal density at the stellar surface $\rho_\star$, and velocities are expressed in units of Keplerian velocity at the stellar surface $v_{\text{K},\star} = (GM_\star/R_\star)^{1/2}$, given the stellar mass $M_\star$. Hence, time is expressed in units of $t_0 = R_\star / v_{\text{K},\star}$. Magnetic field strength is expressed in units of $B_0 = (4 \pi \rho_\star v_{\text{K},\star}^2)^{1/2}$. Mass accretion/ejection rates are expressed in units of $\dot{M}_0 = \rho_\star R_\star^2 v_{\text{K},\star}$, and torque is expressed in units of $\dot{J}_0 = \rho_\star R_\star^3 v_{\text{K},\star}^2$.

In order to make direct comparisons with young stars, we adopt $R_\star = 2$ $R_\odot$, $M_\star = 0.5$ $M_\odot$, and $\rho_\star = 10^{-12}$ g cm$^{-3}$ and express the normalizations as

\begin{align}\label{eq:norm}
\begin{gathered}
v_{\text{K},\star}  = 218.38  \left(\tfrac{M_\star}{0.5 \, M_\odot}\right)^{1/2}  \left(\tfrac{R_\star}{2 \, R_\odot}\right)^{-1/2}  \,  \text{km s}^{-1}\\
B_0  = 77.41 \left(\tfrac{\rho_\star}{10^{-12} \, \text{g} \, \text{cm}^{-3}}\right)^{1/2}  \left(\tfrac{M_\star}{0.5 \, M_\odot}\right)^{1/2}  \left(\tfrac{R_\star}{2 \, R_\odot}\right)^{-1/2} \,  \text{G} \\
t_0 =  0.074 \left(\tfrac{M_\star}{0.5 \, M_\odot}\right)^{-1/2} \left(\tfrac{R_\star}{2 \, R_\odot}\right)^{3/2} \, \text{days} \\
\dot{M}_0 =  6.71 \times 10^{-9} \left(\tfrac{\rho_\star}{10^{-12} \, \text{g cm}^{-3}}\right) \left(\tfrac{M_\star}{0.5 \, M_\odot}\right)^{1/2} \left(\tfrac{R_\star}{2 \, R_\odot}\right)^{3/2} \, M_\odot \, \text{yr}^{-1} \\
\dot{J}_0 =  1.29 \times 10^{36} \left(\tfrac{\rho_\star}{10^{-12} \, \text{g cm}^{-3}}\right) \left(\tfrac{M_\star}{0.5 \, M_\odot}\right) \left(\tfrac{R_\star}{2 \, R_\odot}\right)^{2} \, \text{erg}.
\end{gathered}
\end{align}

\subsection{Simulation parameters}\label{sec:sim_param}

In our parameter study, we vary the following input parameters:

\begin{enumerate}
\item{The initial disk density, $\rho_{\text{d,}\star}$,} 
\item{Surface polar magnetic field strength $B_\star$, controlled by the parameter $v_\text{A}/v_\text{esc}$, i.e., the ratio of the surface polar Alfvén velocity $v_\text{A} = B_\star / (4 \pi \rho_\star)^{1/2}$ and the stellar escape velocity $v_\text{esc}=(2GM_\star/R_\star)^{1/2}$,}
\item{Stellar break-up fraction $f=\Omega_\star R_\star / v_{\text{K},\star}$, where $\Omega_\star$ is the stellar rotation rate.}
\end{enumerate}
The stellar rotation period can be written as

\begin{equation}\label{eq:period}
P_\star = 9.30 \left(\frac{f}{0.05}\right)^{-1} \left(\frac{M_\star}{0.5 \, M_\odot}\right)^{-1/2} \left(\frac{R_\star}{2 \, R_\odot}\right)^{3/2} \, \text{days}.
\end{equation}
We fix the disk thermal aspect-ratio at $\epsilon=0.075$, the viscous and resistive transport coefficients at $\alpha_\text{v}=0.2$ and $\alpha_\text{m}=0.2$, respectively, and the stellar wind sound speed at the stellar surface at $c_{\text{s},\star} = 0.35 v_{\text{K},\star}$. The initial accretion rate of the disk, determined solely by the viscous torque, is

\begin{equation}\label{eq:mass_acc_i}
\dot{M}_\text{acc,i} \approx 0.12 \left(\frac{\alpha_\text{v}}{0.2}\right) \left(\frac{\rho_{\text{d,}\star}}{100}\right) \left(\frac{\epsilon}{0.075}\right)^{3} \, \dot{M}_0.
\end{equation}

\section{Simulations of star-disk interaction}\label{sec:sims_chapter}

We run 38 simulations in total, changing the simulation parameters described in Section~\ref{sec:sim_param}. {Each simulation is run} for a period that corresponds to roughly 20 stellar rotation periods for a break-up fraction $f=0.05$ ($t_0=2513.2742$), therefore less stellar rotation periods are covered as $f$ decreases. 
However, we still find sensible time-averages over a quasi-steady state for our lowest $f$ simulations.
{Typical simulation times in our study correspond to a period of the order of weeks or months, where changes in the stellar rotation rate would be negligible, thus using a fixed $f$ for each simulation is justified.}
The input parameters and  outputted variables (in simulation units) for all simulations can be found in Table~\ref{tab:Pluto_sims}; for convenience, normalization factors for converting these parameters to physical units are shown in Table~\ref{tab:conversion} (see Section~\ref{sec:units_norm}). {Negative mass fluxes or torques represent an inflow of mass or angular momentum onto the star, whereas positive quantities represent an outflow from the star.}

\begin{deluxetable*}{ccccc|cccccccccccccc}
\tabletypesize{\footnotesize}

\tablecaption{Models and variable input parameters (left of vertical line), and outputted global variables (right).\label{tab:Pluto_sims}}

\tablehead{\rule{0pt}{4ex} Model & $f$ & $\frac{\rho_{\text{d,}\star}}{\rho_\star}$ & $\frac{v_\text{A}}{v_\text{esc}}$ & $\frac{B_\star}{B_0}$$^1$ & $\frac{\dot{M}_\text{wind}}{\dot{M}_0}$ & $\frac{\dot{M}_{\text{ME},\star}}{\dot{M}_0}$ & $\frac{\dot{M}_\text{acc}}{\dot{M}_0}$ & $\frac{\dot{J}_\text{wind}}{\dot{J}_0}$ & $\frac{\dot{J}_{\text{ME},\star}}{\dot{J}_0}$ & $\frac{\dot{J}_\text{acc}}{\dot{J}_0}$ & $\frac{\Phi_\text{wind}}{\Phi_0}$ & $\frac{\Phi_{\text{ME},\star}}{\Phi_0}$ & $\frac{\Phi_\text{acc}}{\Phi_0}$ & $\frac{\langle r_\text{A} \rangle}{R_\star}$ & $\frac{R_\text{t}}{R_\star}$ & $\Upsilon_\text{wind}$ & $\Upsilon_\text{acc}$ \\
 & & & & & [$10^{-2}$] & [$10^{-2}$] & & [$10^{-1}$] &  &  & & & & & &  [$10^4$] & }

\startdata
1 & 0.001  & 100 & 4.60 & 12.8 & 0.308 & -0.131 & -0.327 & 0.0381 & -0.194 & -0.438 & 6.77 & 4.81 & 69.6 & 35.04 & 5.47 & 1.05 & 352  \\
2& 0.001  & 200 & 4.60 & 12.8 & 0.401 & -0.744 & -0.627 & 0.0643 & -0.329 & -0.814 & 9.17 & 5.99 & 66.0 & 40.0 & 4.46 & 1.48 & 184  \\
3& 0.001  & 300 & 4.60 & 12.8 & 0.438 & -0.952 & -1.03  & 0.0946 & -0.541 & -1.21 & 11.7 & 7.03 & 62.4 & 46.5 & 3.78 & 2.23 & 112  \\
4& 0.001  & 400 & 4.60 & 12.8 & 0.456 & -1.61  & -1.48  & 0.128  & -0.730 & -1.68 & 14.1 & 7.99 & 59.1 & 52.6 & 3.33 & 3.06 & 78.1  \\
5& 0.0025 & 100 & 4.60 & 12.8 & 0.368 & -0.177 & -0.334 & 0.126  & -0.211 & -0.451 & 7.92 & 3.85 & 69.4 & 37.0 & 5.42 & 1.21 & 345  \\
6& 0.0025 & 200 & 4.60 & 12.8 & 0.462 & -0.841 & -0.625 & 0.206 & -0.304 & -0.820 & 10.4 & 4.66 & 66.1 & 42.1 & 4.47 & 1.67 & 184  \\
7& 0.0025 & 300 & 4.60 & 12.8 & 0.495 & -0.986 & -1.09  & 0.315 & -0.526 & -1.29 & 13.5 & 5.65 & 62.0 & 50.4 & 3.73 & 2.60 & 106  \\
8& 0.0025 & 400 & 4.60 & 12.8 & 0.517 & -1.96  & -1.46  & 0.398 & -0.729 & -1.67 & 15.8 & 6.37 & 59.0 & 55.5 & 3.33 & 3.43 & 78.7  \\
9& 0.005  & 100 & 4.60 & 12.8 & 0.395 & -0.160 & -0.38  & 0.297 & -0.214 & -0.498 & 8.44 & 3.64 & 69.1 &  38.8 & 5.21 & 1.27 & 303  \\
10& 0.005  & 200 & 4.60 & 12.8 & 0.497 & -0.628 & -0.711 & 0.51  & -0.326 & -0.909 & 11.5 & 4.31 & 65.4 & 45.3 & 4.31 & 1.89 & 162  \\
11& 0.005  & 300 & 2.30 & 6.38 & 0.521 & -4.55  & -1.30  & 0.711 & -0.643 & -1.26 & 14.3 & 4.49 & 21.8 & 52.3 & 2.12 & 2.77 & 22.1  \\
12& 0.005  & 300 & 4.60 & 12.8 & 0.533 & -1.11  & -1.11  & 0.737 & -0.510 & -1.33 & 14.5 & 5.00 & 61.7 & 52.6 & 3.72 & 2.77 & 103  \\
13& 0.005  & 300 & 9.20 & 25.5 & 0.365 & -0.300 & -1.06  & 0.579 & -0.569 & -1.58 & 15.1 & 6.44 & 141 & 56.4 & 5.89 & 4.42 & 434  \\ 
14& 0.005  & 400 & 4.60 & 12.8 & 0.559 & -1.90  & -1.56  & 0.939 & -0.697 & -1.79 & 17.2 & 5.51 & 58.5 & 58.0 & 3.29 & 3.73 & 73.8  \\
15& 0.01   & 100 & 4.60 & 12.8 & 0.425 & -0.434 & -0.380 & 0.663 & -0.197 & -0.533 & 8.99 & 3.44 & 68.8 & 39.5 & 5.19 & 1.34 & 303  \\
16& 0.01   & 200 & 2.30 & 6.38 & 0.516 & -2.46  & -0.810 & 1.10  & -0.407 & -0.827 & 11.5 & 3.34 & 25.8 & 46.1 & 2.54 & 1.80 & 35.5  \\
17& 0.01   & 200 & 4.60 & 12.8 & 0.525 & -0.597 & -0.763 & 1.12  & -0.327 & -0.971 & 12.2 & 4.19 & 64.8 & 46.3 & 4.23 & 2.01 & 151 \\
18& 0.01   & 200 & 9.20 & 25.5 & 0.325 & -0.231 & -0.791 & 1.17 & -0.419 & -1.28 & 13.5 & 5.55 & 143 & 59.9 & 6.40 & 3.93 & 582  \\
19& 0.01   & 300 & 4.60 & 12.8 & 0.559 & -1.22  & -1.21  & 1.58 & -0.507 & -1.45 & 15.3 & 4.73 & 61.1 & 53.2 & 3.61 & 2.97 & 95.5  \\
20& 0.01   & 400 & 4.60 & 12.8 & 0.580 & -2.20  & -1.61  & 2.09 & -0.651 & -1.84 & 17.8 & 5.18 & 58.2 & 60.0 & 3.25 & 3.87 & 71.5  \\
21& 0.025  & 100 & 4.60 & 12.8 & 0.535 & -0.302 & -0.583 & 1.96 & -0.210 & -0.797 & 11.1 & 3.53 & 66.6 & 38.3 & 4.64 & 1.62 & 198  \\
22& 0.025  & 200 & 4.60 & 12.8 & 0.610 & -0.985 & -1.09  & 3.18 & -0.409 & -1.32 & 14.9 & 4.39 & 61.9 & 45.5 & 3.74 & 2.56 & 105  \\
23& 0.025  & 300 & 4.60 & 12.8 & 0.647 & -1.55  & -1.51  & 4.27  & -0.547 & -1.79 & 17.8 & 4.75 & 58.7 & 51.4 & 3.34 & 3.45 & 76.5  \\
24& 0.025  & 300 & 9.20 & 25.5 & 0.477 & -0.759 & -1.80  & 4.49  & -0.724 & -2.61 & 19.9 & 7.19 & 135 & 61.3 & 5.01 & 5.85 & 256  \\
25& 0.025  & 400 & 3.45 & 9.57 & 0.669 & -5.32  & -2.02  & 5.22  & -0.795 & -2.16 & 20.2 & 4.67 & 36.1 & 55.9 & 2.41 & 4.29 & 32.1  \\
26& 0.025  & 400 & 4.60 & 12.8 & 0.676 & -2.46  & -1.91  & 5.44  & -0.687 & -2.22 & 20.4 & 5.09 & 55.7 & 56.7 & 3.03 & 4.33 & 60.2  \\
27& 0.05   & 100 & 2.30 & 6.38 & 0.822 & -1.40  & -0.656 & 3.56 & -0.216 & -0.712 & 11.5 & 2.46 & 26.6 & 29.4 & 2.73 & 1.14 & 43.9  \\
28& 0.05   & 100 & 3.45 & 9.57 & 0.814 & -0.802 & -0.787 & 4.49 & -0.206 & -0.973 & 13.1 & 3.10 & 44.7 & 33.2 & 3.41 & 1.49 & 82.3  \\
29& 0.05   & 100 & 4.60 & 12.8 & 0.774 & -0.499 & -0.955 & 5.50 & -0.244 & -1.28 & 15.0 & 3.91 & 62.2 & 37.7 & 3.87 & 2.07 & 121  \\
30& 0.05   & 200 & 3.45 & 9.57 & 0.848 & -2.15  & -1.34  & 6.68  & -0.401 & -1.50 & 16.8 & 3.67 & 40.5 & 39.7 & 2.83 & 2.34 & 48.4  \\
31& 0.05   & 200 & 4.60 & 12.8 & 0.837 & -1.27  & -1.55  & 7.83  & -0.436 & -1.88 & 18.5 & 4.48 & 58.2 & 43.3 & 3.32 & 2.88 & 74.4  \\
32& 0.05   & 300 & 3.45 & 9.57 & 0.897 & -4.76  & -2.01  & 9.15  & -0.666 & -2.20 & 20.7 & 4.11 & 36.0 & 45.2 & 2.41 & 3.39 & 32.2  \\
33& 0.05   & 300 & 4.60 & 12.8 & 0.880 & -2.28  & -2.13  & 9.97  & -0.625 & -2.45 & 21.6 & 4.81 & 54.8 & 47.5 & 2.96 & 3.75 & 54.1  \\
34& 0.05   & 400 & 3.45 & 9.57 & 0.943 & -8.19  & -3.21  & 12.1  & -1.11 & -3.30 & 24.9 & 4.97 & 31.0 & 50.7 & 2.09 & 4.64 & 20.2  \\
35& 0.05   & 400 & 4.60 & 12.8 & 0.931 & -4.34  & -2.96  & 12.6  & -0.867 & -3.32 & 25.4 & 5.45 & 50.4 & 52.0 & 2.60 & 4.89 & 39.0  \\
36& 0.0625 & 200 & 4.60 & 12.8 & 0.999 & -1.82  & -1.79  & 10.4  & -0.447 & -2.17 & 20.2 & 4.79 & 56.2 & 40.9 & 3.12 & 2.87 & 64.5  \\
37& 0.0625 & 300 & 4.60 & 12.8 & 1.02  & -2.49  & -2.35  & 12.6  & -0.629 & -2.71 & 22.7 & 5.03 & 53.4 & 44.4 & 2.84 & 3.56 & 48.9  \\
38& 0.0625 & 400 & 4.60 & 12.8 & 1.09  & -5.56  & -3.40  & 16.0  & -0.868 & -3.79 & 26.8 & 6.00 & 48.3 & 48.3 & 2.46 & 4.66 & 33.9  \\
\enddata

\tablenotetext{1}{$B_\star$ is not a fundamental input parameter, but it is simply derived from $v_\text{A}/v_\text{esc}$ and tabulated here for convenience.}

\end{deluxetable*}

\begin{deluxetable}{cc}
\tablecaption{Normalization of each parameter type in Table~\ref{tab:Pluto_sims}.\label{tab:conversion}}

\tablehead{Parameter & Normalization}

\startdata
$B_0$ & $(4 \pi \rho_\star v_{\text{K},\star}^2)^{1/2}$ \\
$\dot{M}_0$ & $\rho_\star R_\star^2 v_{\text{K},\star}$ \\
$\dot{J}_0$ & $\rho_\star R_\star^3 v_{\text{K},\star}^2$ \\ 
$\Phi_0$ & $(4 \pi \rho_\star R_\star^4 v_{\text{K},\star}^2)^{1/2}$ \\
\enddata
\end{deluxetable}

\subsection{Morphology of star-disk interaction systems}\label{sec:morph}

In this section, we qualitatively analyze the properties of the SDI system, focusing on the dynamical processes that are a result of this interaction, namely, the stellar wind, the accretion flow, and the time-dependent magnetospheric ejections (MEs). 
\citet{refId0} have a detailed analysis of this latter phenomenon, which we briefly describe below. Due to the time-dependent nature of these simulations, they can only achieve a quasi-steady or quasi-periodic state; however, as the focus of this paper is to investigate global properties and the long-term evolution of such a system, any periodic transient effects have been neglected by using time-averages.

\begin{figure}
\begin{center}
\includegraphics[width=0.4625\textwidth]{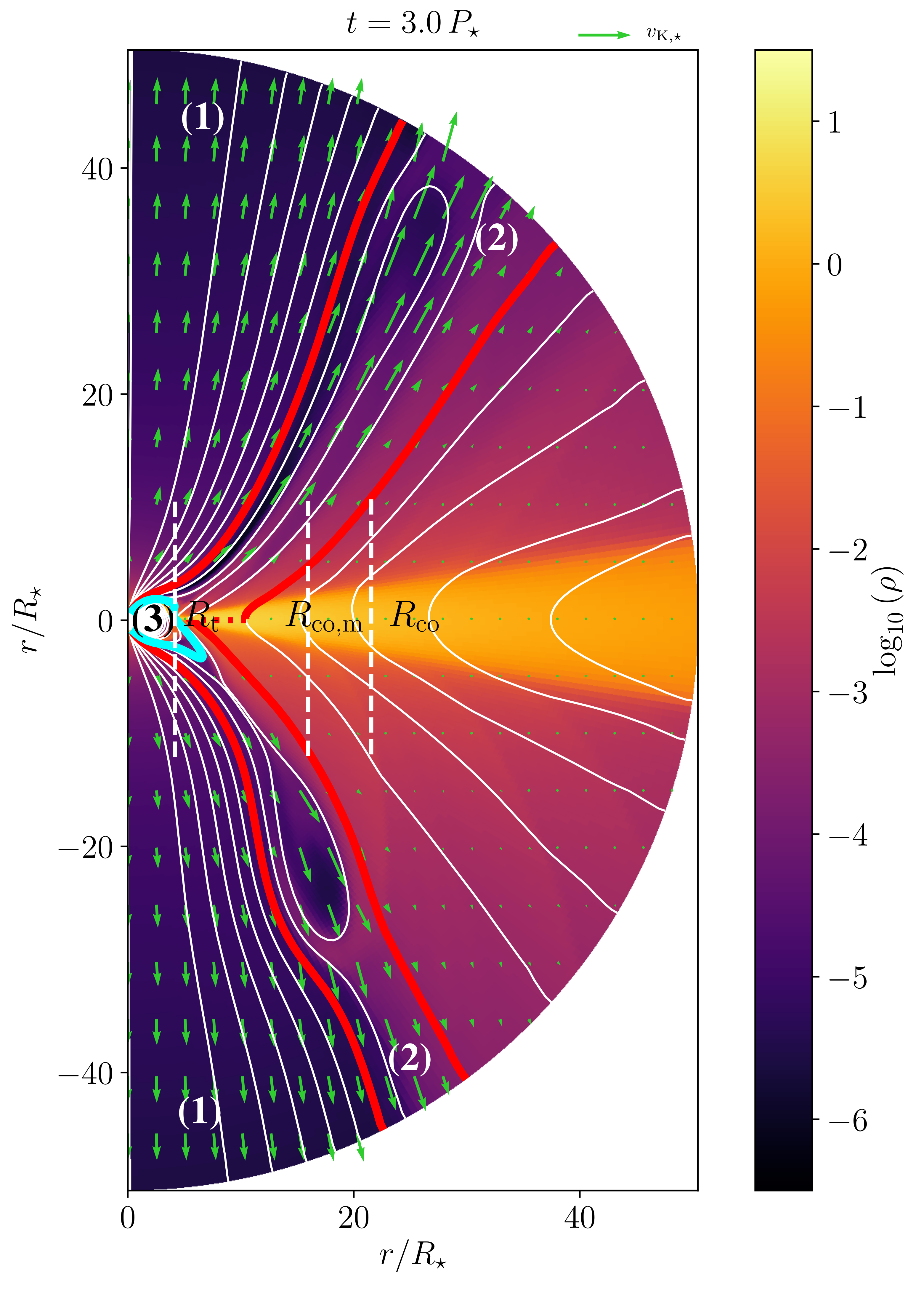}
\caption{A logarithmic density color map showing the entire domain of model 17, at $t = 3 P_\star$, where $P_\star$ is the stellar rotation period. Magnetic field lines (white) and velocity vectors normalized by $v_\text{K},\star$ (green arrows) are included. We illustrate the three regions that contribute to the net stellar torque: (1) the stellar wind, (2) MEs, and (3) accretion flow. Region (2) is bounded by the red field lines, and regions (2) and (3) are separated by the cyan field line. Vertical white dashed lines (from left to right) show the position of the truncation radius, $R_\text{t}$, the magnetic corotation radius, $R_\text{co,m}$, and the Keplerian corotation radius, $R_\text{co}$, respectively.  A zoomed-in region of this plot is shown in Figure~\ref{fig:density_zoom}. \label{fig:density_150}}
\end{center}
\end{figure}

\begin{figure*}
\begin{center}
\includegraphics[width=0.725\textwidth]{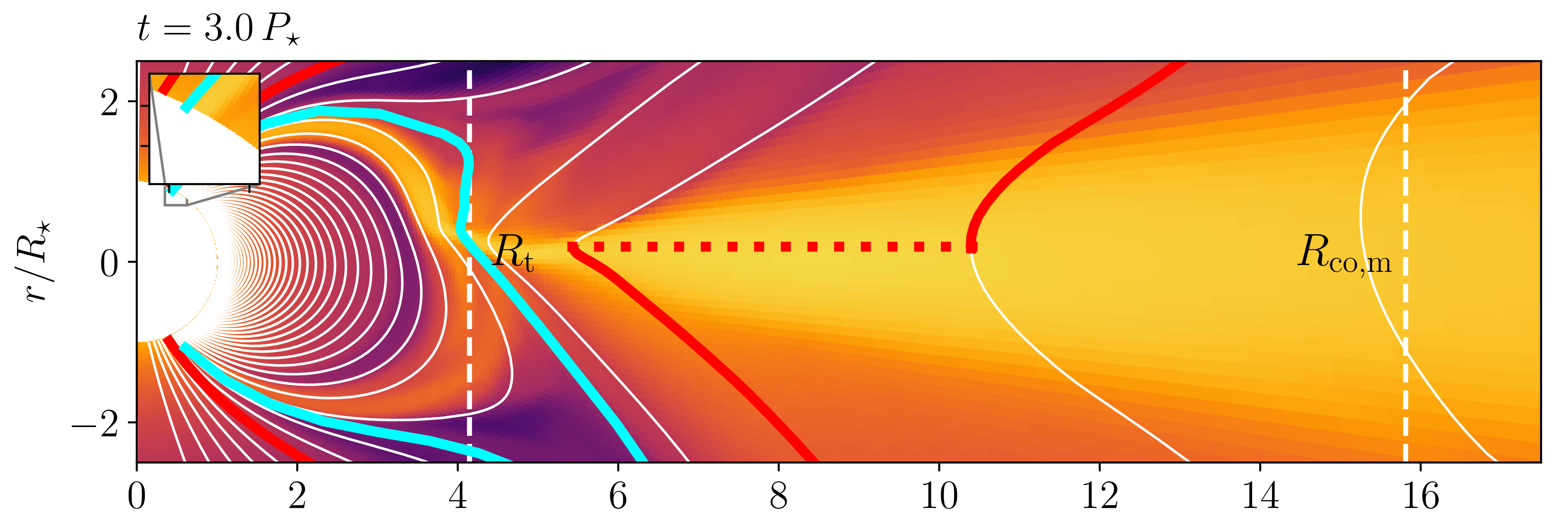}{(a)}
\vspace{1mm}
\includegraphics[width=0.725\textwidth]{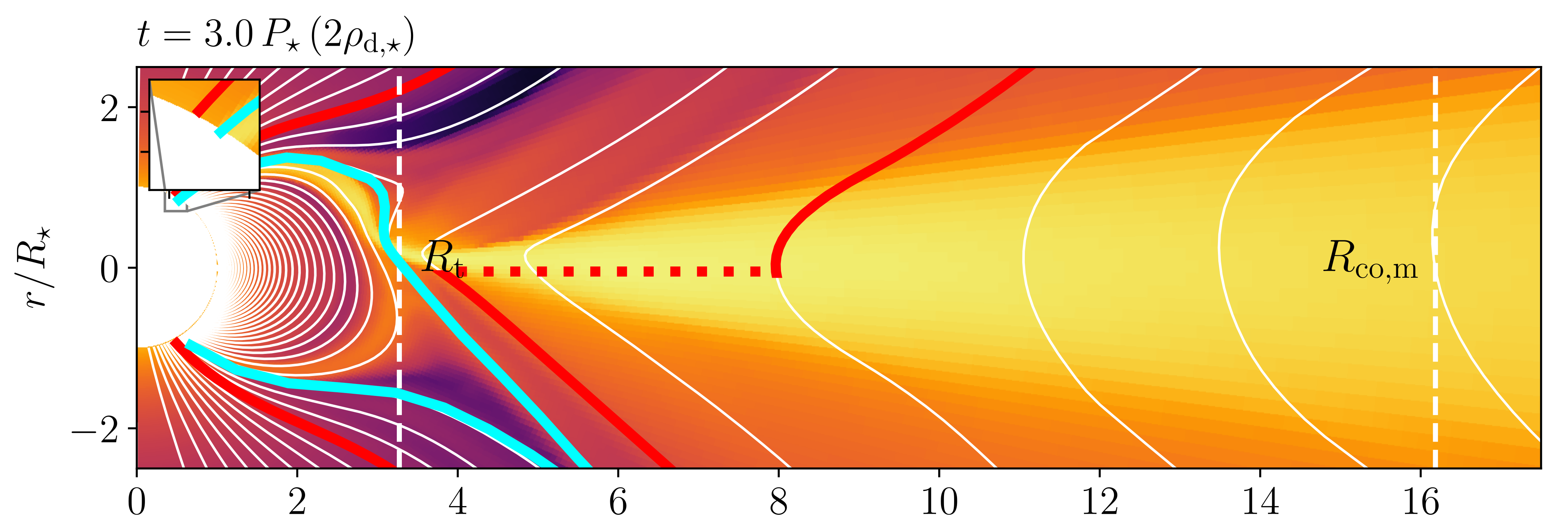}{(b)}
\vspace{1mm}
\includegraphics[width=0.725\textwidth]{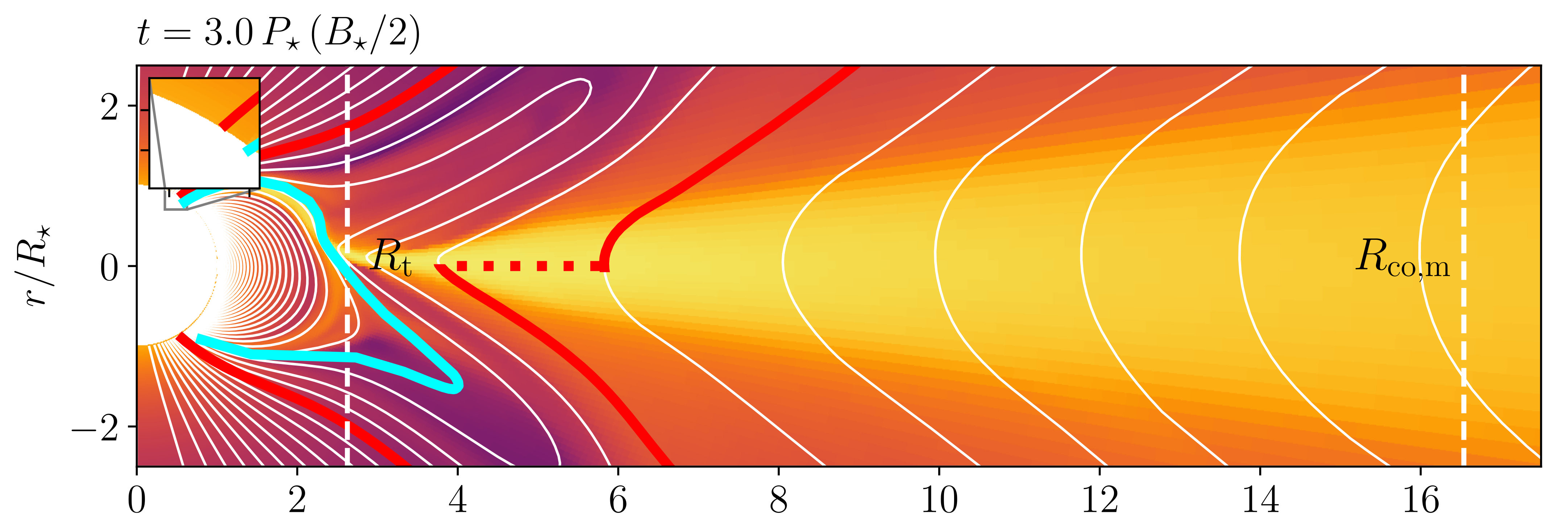}{(c)}
\vspace{1mm}
\includegraphics[width=0.725\textwidth]{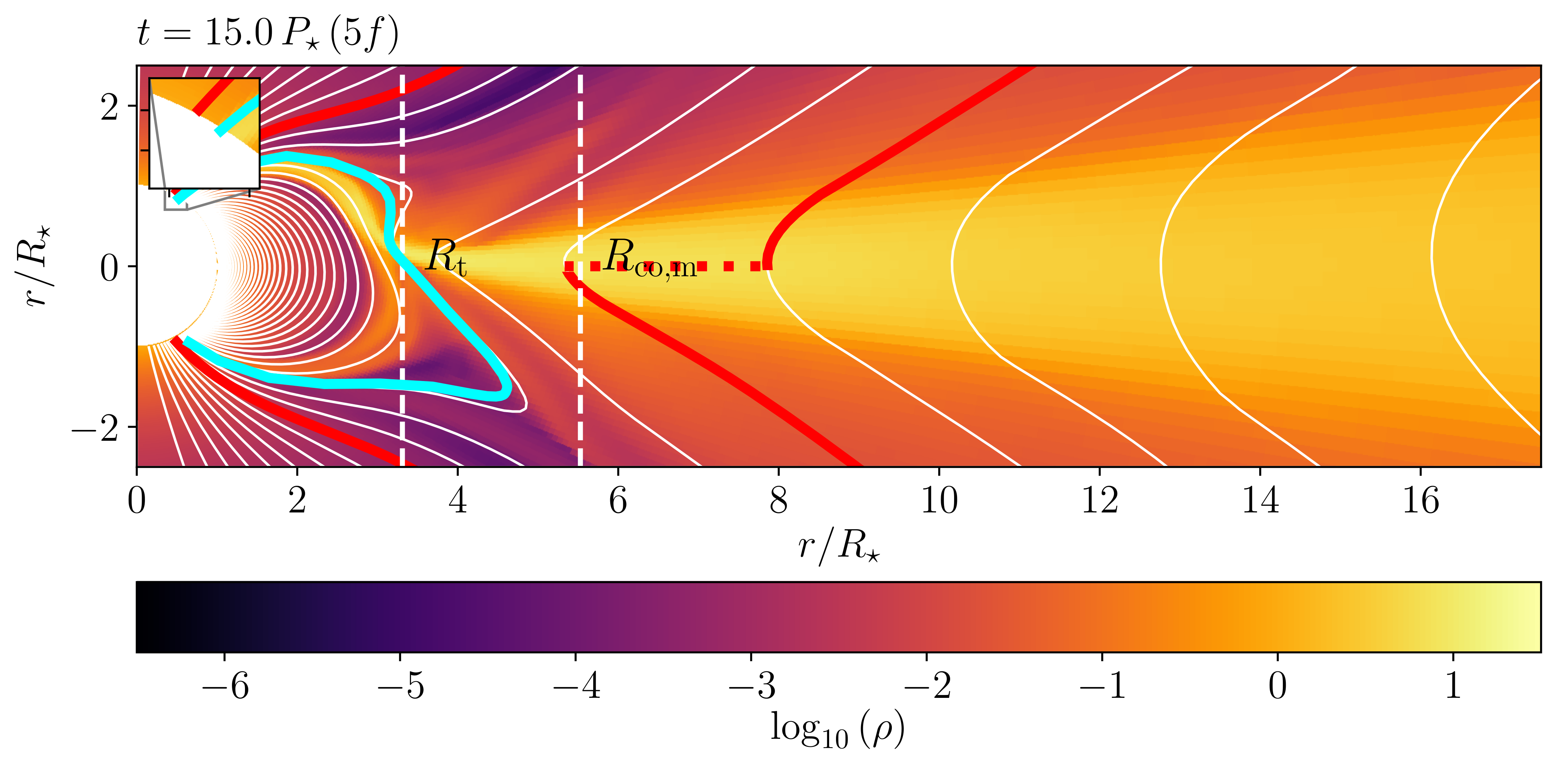}{(d)}
\caption{A logarithmic density color map showing the inner domain of (a) model 17 (fiducial), (b) model 20 (twice the disk density), (c) model 16 (half the magnetic field strength), and (d) model 31 (five times the rotation rate). For each simulation, magnetic field lines (white) are included. The stellar wind and ME regions are separated by the red field lines, and the ME and accretion regions are separated by the cyan field line. Vertical white dashed lines (from left to right) show the position of the truncation radius, $R_\text{t}$, and the magnetic corotation radius, $R_\text{co,m}$, respectively. \label{fig:density_zoom}}
\end{center}
\end{figure*}

Figure~\ref{fig:density_150} shows the density distribution for the computational domain of a representative SDI simulation (model 17), taken at a time corresponding to 3 stellar rotation periods (middle of time-averaged domain).  
The top panel of Figure~\ref{fig:density_zoom} also shows a zoomed-in region of this Figure~\ref{fig:density_150}.  
The magnetic field (white) is strongly coupled to the accretion disk. 
We dissect the system into three distinct regions.
Region (1) shows the ejection of the stellar wind occurring along these open magnetic field lines anchored to the stellar surface, providing the star with a spin-down torque contribution. 

Region (2) contains the MEs, exhibiting the inflation and reconnection of magnetospheric field lines. These occur due to the differential rotation between the star and the disk, resulting in a strong toroidal field that inflates the magnetosphere and opens field lines close to the truncation radius (labeled $R_\text{t}$). This causes ejections to detach from the magnetosphere and propagate outwards in between the open magnetic field lines and those connected to the disk, bounded by the red field lines (a dotted red line is used to connect asymmetric ME regions about the equatorial plane, and should not be interpreted as a magnetic field line). 

At the inner disk edge, the magnetic field lines connect the disk to the star, and act as accretion funnels that allow the exchange of angular momentum between the disk and the stellar surface to occur at the truncation radius, which is shown in region (3). Region (3) also contains the ``dead-zone" region, which consists of corotating {coronal gas below the accretion flow}, around the magnetic equator. 
The incoming angular momentum from accretion provides the star with a spin-up torque contribution.  Although our system is initialized with north-south symmetry (other than the magnetic polarity reversal, characteristic of a dipole field), our simulation setup does not assume or enforce any north-south symmetry during the evolution. We observe some models to establish funnel flows that are approximately symmetric, some to be dominated in the north or south hemispheric flows, and some that switch between regimes one or more times during the simulation. This can also produce hemispheric asymmetry in the MEs. For our example in Figure~\ref{fig:density_150}, the dominant accretion channel occurs in the northern hemisphere. The determination of mass loss rates, torques, and unsigned magnetic fluxes associated with each of these mechanisms is discussed in Section~\ref{sec:global_flow}.

Our criterion for defining the boundary between regions (1) and (2) selects the magnetic field lines closest to the midplane (one for each hemisphere) that is attached to the star and reaches the outer domain, and where both $v_R > 0$ (ejection) and $B_R B_\phi < 0$ (spin-down torque). These field lines also correspond to the boundary between region (2) and the outer disk domain. Our criterion for defining the boundary between regions (2) and (3) selects the outermost closed magnetic field line that has a maximum $R$ coordinate no more than $5 \%$ larger than its midplane radius in at least one of the hemispheres (in Figure~\ref{fig:density_150}, this criteria is met in the upper hemisphere, where the dominant accretion channel occurs), and where flow is supersonic at its connection with the stellar surface.
This avoids misclassifying field lines that are still closed but are strongly deformed due to MEs (and in practice, these strongly deformed field lines open and close intermittently, while the lines we select typically remain closed for the duration). 
The truncation radius is then taken to be the radial coordinate corresponding to the minimum specific entropy (which locates the densest region of the disk material) along this last closed field line \citep{2020arXiv200900940P}.

We plot the Keplerian corotation radius $R_\text{co}=f^{-2/3} R_\star$, which represents the radius at which a Keplerian disk rotates at the same speed as the star. 
For such a disk, if $r<R_\text{co}$, the magnetic field is twisted such that field lines ``lead" the stellar rotation, thus magnetic torques threading this region spin up the star; conversely, magnetic torques threading regions where $r>R_\text{co}$ spin down the star. 
However, MEs extract angular momentum from the accretion disk, resulting in both a smaller accretion torque being exerted onto the stellar surface and the disk rotating slower than Keplerian speed. Thus in general, for $r<R_\text{co}$, the star-disk differential rotation is reduced, resulting in a weaker magnetic ``twisting" effect and reduced magnetic torques, relative to Keplerian rotation. 
Thus, we plot the magnetic corotation radius, $R_\text{co,m}$, defined such that the magnetic twist becomes zero for $R_\text{t}=R_\text{co,m}$. 
The technical criteria of how we determine $R_\text{co,m}$ in our simulations is discussed in Section~\ref{sec:acc_torque}.

$R_\text{t}$ must be less than $R_\text{co}$ for disk material to lose angular momentum and accrete onto the star \citep{2005MNRAS.356..167M}; for a disk truncated outside $R_\text{co}$, i.e., in the ``propeller" regime, accretion is not possible \citep[see, e.g.,][]{1975A&A....39..185I} or intermittent \citep[see, e.g.,][]{Romanova_2005,2010MNRAS.406.1208D}. How the MEs exchange angular momentum with the star depends on the star-disk differential rotation in the truncation region. 
For a system where $R_\text{t} \gtrsim R_\text{co,m}$, the MEs extract angular momentum from a region of the disk rotating slower than the star, and therefore the MEs exert a spin-down torque contribution on the star. 
However, all of our simulations reside in the regime where $R_\text{t} \lesssim R_\text{co,m}$, so the MEs extract angular momentum from a region of the disk rotating faster than the star, and we observe that the MEs exert a time-averaged spin-up torque contribution on the star. 
The region $R_\text{co,m} \lesssim R_\text{t} \lesssim R_\text{co}$ corresponds to a situation where it is possible that the MEs exert a spin-down torque on the stellar surface whilst the disk is still sub-Keplerian, thus in principle can still be steadily accreting.

In Figure~\ref{fig:density_zoom}, we plot a density distribution of the inner domain for four SDI models, taken at a time corresponding to the middle of their time-averaged domain, to demonstrate the qualitative effects of changing the initial disk density, $\rho_{\text{d,}\star}$, the magnetic field strength, $B_\star$, and the stellar break-up fraction, $f$. These include (a) our representative model (model 17), (b) a disk with double the initial disk density (model 20), (c) a star with half the magnetic field strength (model 16), and (d) a star rotating 5 times faster (model 31). 
All three latter cases essentially increase the ratio between the disk pressure and magnetic stress, which decreases $R_\text{t}$ compared to our fiducial model. 
Generally, increasing $\rho_{\text{d,}\star}$ increases the mass accretion rate and hence the disk pressure, increasing $f$ decreases the differential rotation between the star and the disk, decreasing the magnetic stress felt by the disk, and decreasing $B_\star$ directly decreases the magnetic stress. $R_\text{co,m}$ decreases with $f$, but also has a small inverse proportionality on $R_\text{t}$; we demonstrate this quantitatively in Section~\ref{sec:acc_torque}. For all cases, we include a zoom panel at identical coordinates, to demonstrate the relative positions of the field lines where accretion and ME mechanisms deposit angular momentum (cyan and red lines, respectively). For cases (b)-(d), a smaller truncation radius results in these field lines connecting to the star at lower latitudes, opening up a larger area on the stellar surface for the stellar wind to be ejected.

\newpage
\subsection{Global flow quantities and efficiencies}\label{sec:global_flow}

The main aim of this paper is to investigate how the stellar wind, accretion flow, and MEs interact with the star.
Thus we investigate global flow properties, such as the mass flow rate, torque, and unsigned magnetic flux. 
Defining a surface $S$ perpendicular to the poloidal flow at a given radius, these quantities are written as

\begin{equation}\label{eq:mass_rate_int}
\dot{M} = \int_{S} \rho \boldsymbol{v_\text{p}} \cdot \text{d}\boldsymbol{S} = 2\pi R^2 \int_{\theta_1}^{\theta_2} \rho v_R \sin{\theta} \, \text{d}\theta,
\end{equation}

\begin{align}\label{eq:torque_int}
\begin{split}
\dot{J} = {}&
\int_{S} r \left(\rho v_\phi \boldsymbol{v_\text{p}} - \frac{B_\phi \boldsymbol{B_\text{p}}}{4 \pi} \right) \cdot \text{d}\boldsymbol{S} \\
 = {}& 2 \pi R^3 \int_{\theta_1}^{\theta_2} \left(\rho v_\phi v_\text{R} - \frac{B_\phi B_\text{R}}{4 \pi} \right) \sin^2{\theta} \, \text{d} \theta,
\end{split}
\end{align}
and

\begin{equation}\label{eq:unsigned_flux}
\Phi = \int_{S} \lvert\boldsymbol{B} \cdot \text{d}\boldsymbol{S}\rvert = 2 \pi R^2 \int_{\theta_1}^{\theta_2} \lvert B_R \sin{\theta} \rvert \, \text{d}\theta,
\end{equation}
respectively, where ${v_\text{R}}$ and ${B_\text{R}}$ are the radial velocity and magnetic field strength, respectively, and $B_\phi$ is the toroidal magnetic field strength. These global quantities can be split into additive contributions from the stellar wind, MEs, and accretion, which we hereafter distinguish with subscripts ``wind", ``ME,$\star$", and ``acc"\footnote{Global quantities with the ``acc" subscript also encapsulate the ``dead-zone" region.}, respectively. The two angles $\theta_1$ and $\theta_2$ represent the area enclosing regions that correspond to each component. 

For our simulations, we wish to investigate the effects of these global quantities at the stellar surface, which can be determined, in theory, by evaluating Equations~(\ref{eq:mass_rate_int})-(\ref{eq:unsigned_flux}) at $R=R_\star$. To avoid numerical effects near the inner boundary, we instead take the median value of mass flow rates and torques between the stellar surface and $2/3$ of the truncation radius. However, we are still able to calculate the unsigned stellar magnetic flux by evaluating Equation~(\ref{eq:unsigned_flux}) at the stellar surface, as the radial magnetic field strength component is kept fixed at its initial dipolar configuration. The total unsigned stellar magnetic flux can also be expressed analytically:

\begin{equation}\label{eq:stellar_unsigned_flux}
\Phi_\star = \Phi_\text{wind} + \Phi_\text{acc} + \Phi_{\text{ME},\star} \equiv \alpha \pi R_\star^2 B_\star,
\end{equation}
where $\alpha=2$ for our dipolar surface configuration. 

\begin{figure}
\begin{center}
\includegraphics[width=0.4625\textwidth]{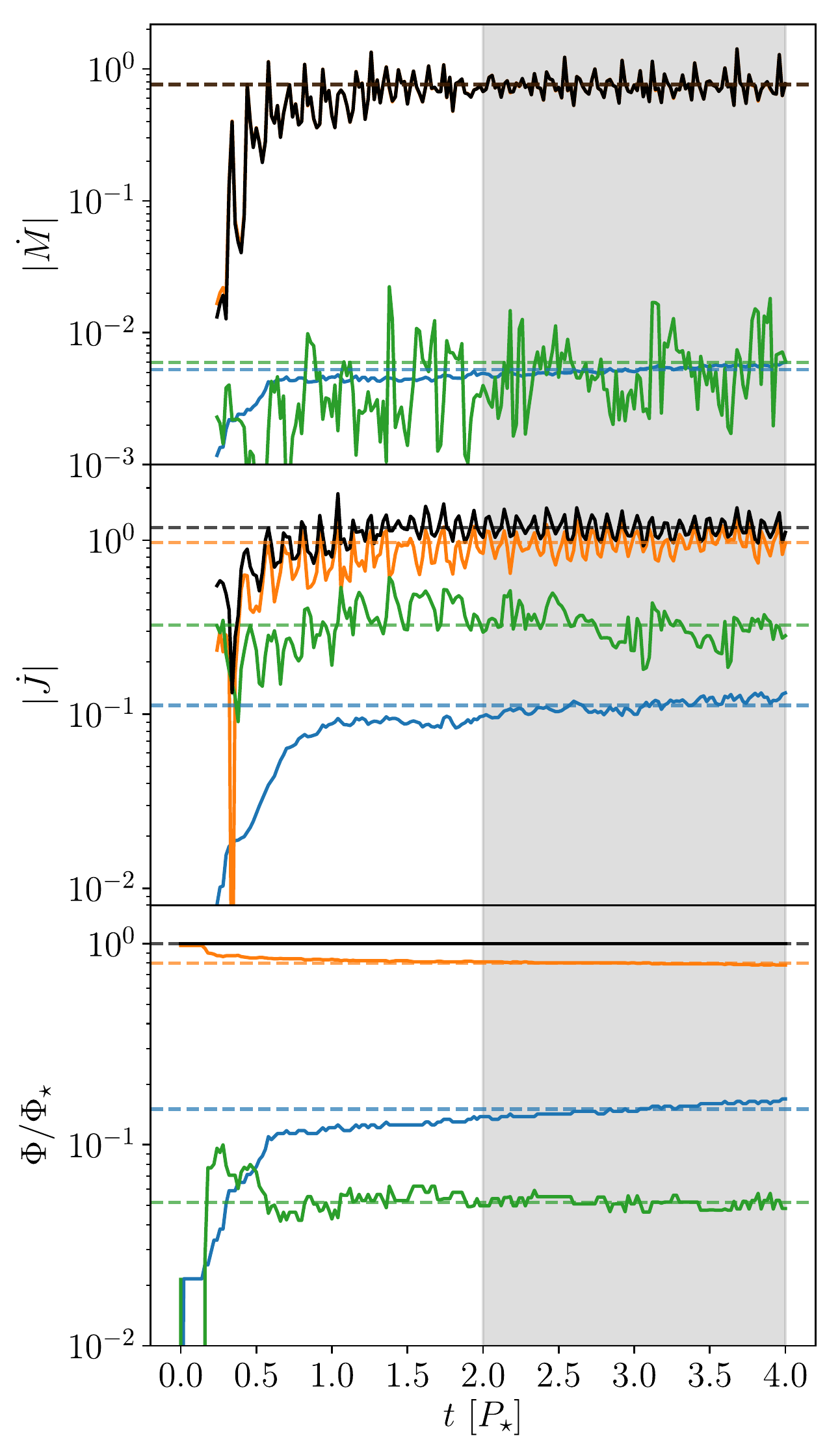}
\caption{Evolution of the stellar (from top to bottom) mass flow rates, torques, and normalized unsigned magnetic fluxes, for the stellar wind (blue), MEs (green), accretion (orange), and the net value (black), as a function of time in units of stellar period, $P_\star$. Respectively colored dotted lines indicate time-averaged values that are taken over a period indicated by the gray shaded region. This figure refers to model 17. \label{fig:global_evo}}
\end{center}
\end{figure}

In Figure~\ref{fig:global_evo}, using model 17 as a representative example, we plot the temporal evolution of the mass flow rates, torques, and unsigned magnetic fluxes (normalized by $\Phi_\star$) for the stellar wind, MEs, accretion, and the sum of these contributions. 
At the beginning of the simulation, there is a large transient event as the simulation adjusts from the initial state to a quasi-steady state, beginning around $t\approx P_*$ for the case shown. 
In this relaxed state, quasi-periodic structure can be seen in the mass flows and torques, which are signatures of the MEs (green).
Furthermore, it is clear that stellar wind (blue) global properties are slowly increasing with time, further demonstrating the quasi-steady configuration of these simulations.
The unsigned accretion magnetic flux (orange) also decreases with time, conserving the unsigned stellar magnetic flux. 
This indicates that the truncation radius is slowly decreasing as the system evolves.
Regardless, we take a time-averaged value for these global quantities (indicated by the gray shaded region) when simulations appear to become quasi-steady state, which we take to be the latter-half of the total timescale for each simulation. 
This method is applied for all simulations, and all values are listed in Table~\ref{tab:Pluto_sims}. 
In spite of the slow drift of the simulated quantities during the period of our time averages, we show in Section~\ref{sec:torque_formulation} that the global properties obey strict relationships and allow us still to derive robust relationships between global system properties.

Our time average mitigates any transient features, such as the periodic behavior of the MEs, and provides global quantities characteristic to the long-term evolution of the system. In the time-averaged domain, a significant proportion of the mass flow rate comes from material accreting onto the stellar surface. {The mass loss rate from the stellar wind and the mass inflow rate from the MEs are comparable.}
A majority of the net stellar torque is also associated with the spin-up contribution from the accretion flow, with a further contribution from the MEs ($\sim 30 \%$), and a small spin-down contribution due to the stellar winds ($\sim 10 \%$); thus this system is in a net spin-up configuration. {In these cases, the MEs accrete some of their material onto the star.}
This hierarchy is also present in the normalized unsigned magnetic flux; however, the high value for accretion is indicative of the inclusion of the dead zone region. Therefore, a majority of the unsigned stellar magnetic flux is closed, and only a small fraction is open in the stellar wind, participating in the intermittent opening and closing of the MEs, or carrying the accretion flow.

\begin{figure}
\begin{center}
\includegraphics[width=0.4625\textwidth]{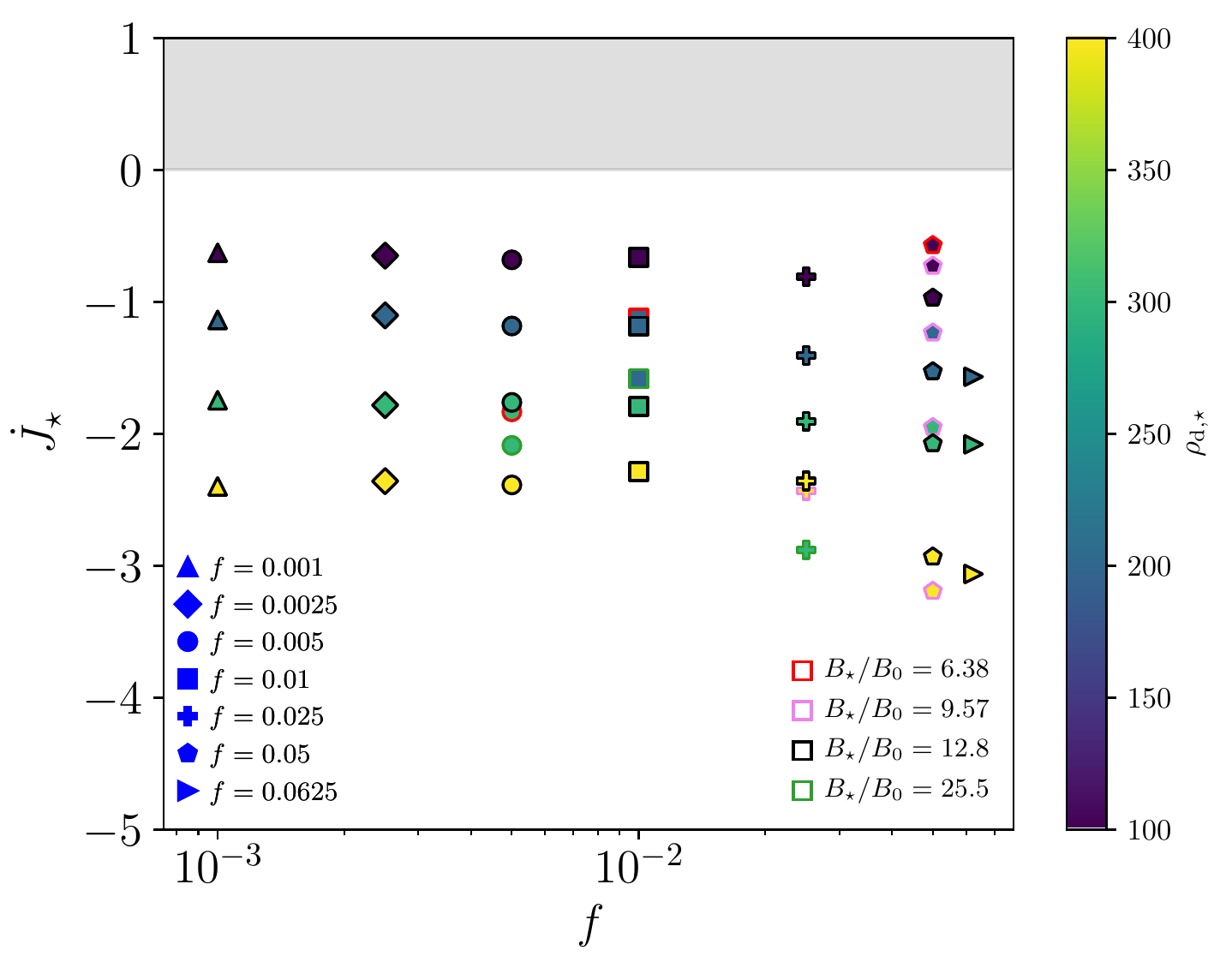}
\caption{The net stellar torque, $J_\star$, as a function of the stellar break-up fraction, $f$, for all simulations. To differentiate between different input parameters, we use differing symbol styles for $f$, marker colors for $\rho_{\text{d,}\star}$, and marker edge colors for $B_\star$. The gray shaded region represents the net spin-down torque regime, whilst the transparent region represents the net spin-up torque regime; all of our simulations exist in the latter. \label{fig:Jdot_star_v_f}}
\end{center}
\end{figure}

To illustrate the extent of the explored parameter space, Figure~\ref{fig:Jdot_star_v_f} shows $\dot{J}_\star$ as a function of $f$ for all simulation, where the symbol shapes, colors, and borders encode the variations of each parameter.
All of our simulations exist in the net spin-up torque regime. 
In general, a decrease in any one of these three input parameters decreases the spin-up torque experienced by the star.
In Section~\ref{sec:torque_formulation}, we show how each component of the net stellar torque depends on these input parameters.

\newpage

\section{Torque formulations for simulations of star-disk interaction} \label{sec:torque_formulation}

\subsection{Disk Truncation Radius} \label{sec:Rt_param}

As disk material spirals inward and gets close enough to the star, the stresses arising from the stellar magnetosphere disrupt fluid motions, resulting in a truncation of the disk at some height above the stellar surface, $R_\text{t}$. In our study, we determine the time-averaged truncation radius over the latter-half of each simulation, where simulations appear to be in a quasi-steady state; these values (normalized by the stellar radius) can be found in Table~\ref{tab:Pluto_sims}.
The truncation radius for a dipolar magnetic field structure is typically parameterized as \citep[see, e.g.,][]{2002ApJ...578..420R, 2005MNRAS.356..167M, 2008A&A...478..155B, 2009A&A...508.1117Z}
\begin{equation} \label{eq:Rt_Y_acc}
	\frac{R_\text{t}}{R_\star} = K_\text{t} \Upsilon_\text{acc}^{m_\text{t}},
\end{equation}
where
\begin{equation} \label{eq:Y_acc}
	\Upsilon_\text{acc} = \frac{B_\star^2 R_\star^2}{4 \pi \lvert \dot{M}_\text{acc} \rvert v_\text{esc}}
\end{equation}
is a ``disk magnetization'' parameter, which effectively represents the ratio of the accretion flow's magnetic field and kinetic energies\footnote{\citet{2020arXiv200900940P} define $B_\star$ as the equatorial magnetic field strength, i.e., $B_{\star{\mathrm{,Pantolmos}}} \equiv B_\star / 2$, therefore $K_\text{t,Pantolmos} = 4^{m_\text{t}} K_\text{t}$.}, $\lvert \dot{M}_\text{acc} \rvert$ is the absolute value of the mass accretion rate, and $v_\text{esc} = (2GM_\star/R_\star)^{1/2}$ is the escape velocity from the stellar surface.\footnote{A full derivation, including both the thermal and ram pressure, would introduce a factor of $[m_\text{s}/(1+m_\text{s}^2)]^{m_\text{t}}$ in Equation~(\ref{eq:Rt_Y_acc}) (as shown in Appendix~\ref{sec:param_Mdot}), where $m_\text{s}$ is the sonic Mach number, which we neglect because we show that a constant $K_\text{t}$ fits the data well.} Values for $\Upsilon_\text{acc}$ in our simulations can be found in Table~\ref{tab:Pluto_sims}.

\begin{figure}
\begin{center}
\includegraphics[width=0.4625\textwidth]{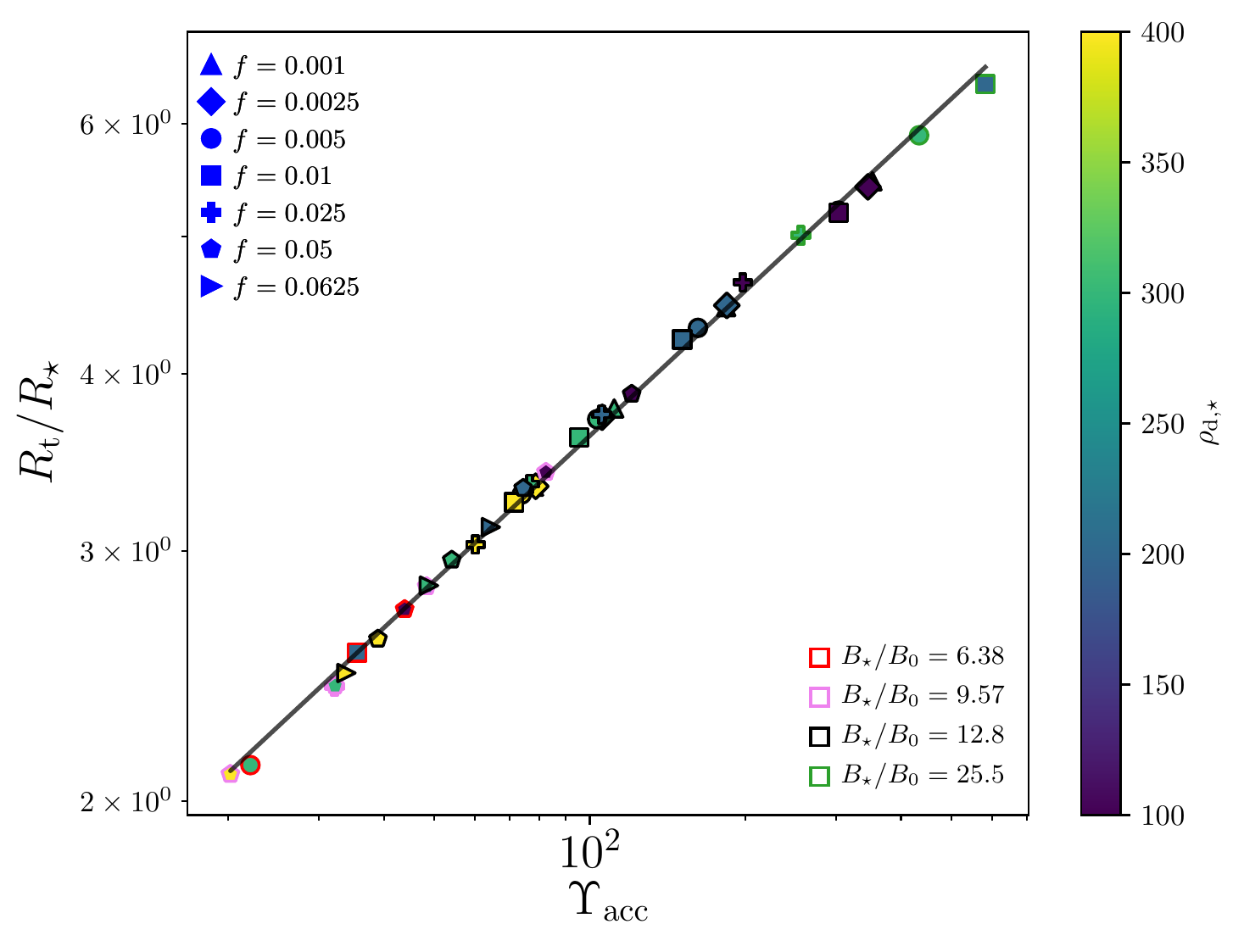}
\caption{Normalized truncation radius, $R_\text{t}/R_\star$, as a function of the disk magnetization parameter, $\Upsilon_\text{acc}$, for all simulations, where the symbols have the same meaning as Figure \ref{fig:Jdot_star_v_f}.  The black line shows the best fit to Equation~(\ref{eq:Rt_Y_acc}), giving $R_\text{t}/R_\star = 0.756 \Upsilon_\text{acc}^{0.340}$. \label{fig:R_t_v_Upsilon_acc}}
\end{center}
\end{figure}

In Figure~\ref{fig:R_t_v_Upsilon_acc}, we plot the normalized truncation radius, $R_\text{t}/R_\star$ as a function of $\Upsilon_\text{acc}$, which shows that all simulations fit the simple power-law formulation of Equation~(\ref{eq:Rt_Y_acc}).
Increasing $B_\star$ results in the disk being truncated further from the star, due to the increased magnetic stresses counteracting the pressure of the disk. 
The truncation radius is also inversely proportional to the mass accretion rate.  
There is no explicit dependence of $R_\text{t}$ on the stellar rotation rate or disk density, but we find that the mass accretion rate in our simulations does depend on these two parameters.  
We present our torque formulations here and throughout in terms of the global mass accretion rate onto the star, so that the formulations are independent of some of the details of our accretion disk structure.  
In Appendix~\ref{sec:param_Mdot}, we show how the accretion rate can be predicted for our particular disk setup, and as a function of the fundamental input parameters.

The best-fit dimensionless parameters to Equation~(\ref{eq:Rt_Y_acc}) are $K_\text{t} = 0.756$ and $m_\text{t} = 0.340$. 
This gives a steeper scaling than the value of $m_\text{t}=2/7$ given by several previous analytic formulations \citep[e.g.,][]{1977ApJ...215..897E, 2002ApJ...578..420R, 2005MNRAS.356..167M, 2008A&A...478..155B}.  Our simulations fit a steeper power-law because (at least in part) the magnetosphere is significantly perturbed relative to the initial, simple scaling of magnetic field strength with $R^{-3}$, which is discussed further in Section~\ref{sec:ME_torque}.
For convenience, all of the best-fit dimensionless parameters presented in Section~\ref{sec:torque_formulation} can be found in Table~\ref{tab:dimensionless_params}.  

\subsection{Accretion torque} \label{sec:acc_torque}

\begin{figure}
\begin{center}
\includegraphics[width=0.4625\textwidth]{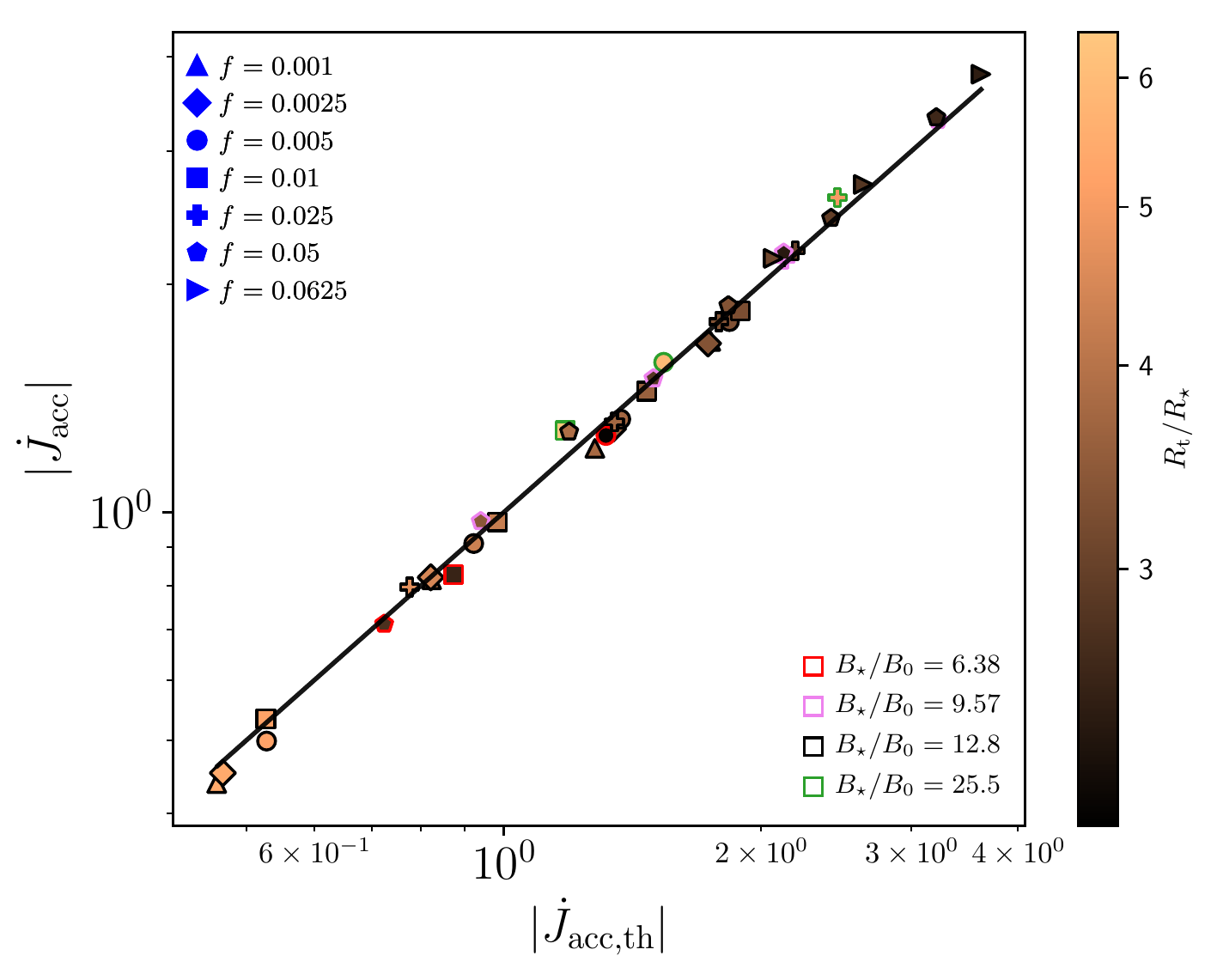}
\caption{The (absolute) accretion torque, $\lvert \dot{J}_\text{acc} \rvert$, as a function of the parameterization in Equation~(\ref{eq:acc_torque_2}) for all simulations. The black line shows $y=x$, illustrating the goodness of fit. The symbols have the same meaning as in Figure~\ref{fig:Jdot_star_v_f}, except that they are colored according to their truncation radius, in this and all following figures. \label{fig:Jdot_acc_with_Kacc_Rt_fitting_function}}
\end{center}
\end{figure}

The accretion torque accounts for the angular momentum carried by the disk material at $R_\text{t}$, which falls onto the star.  
It can be given by
\begin{equation} \label{eq:acc_torque}
	\dot{J}_\text{acc} = - \lvert\dot{M}_\text{acc}\rvert \Omega(R_\text{t}) R_\text{t}^2,
\end{equation}
where $\Omega(R_\text{t})$ is the rotation rate at the truncation radius. The analytical value of the accretion torque can be determined by assuming Keplerian rotation, i.e., $\Omega(R_\text{t})\equiv \Omega_\text{K}(R_\text{t}) = (GM_\star/R_\text{t}^3)^{1/2}$, giving $\dot{J}_\text{acc} = - \lvert\dot{M}_\text{acc}\rvert (GM_\star R_\text{t})^{1/2}$. 
However, \citet{refId0} noted that the accretion disk is not in Keplerian rotation, due to the MEs extracting a fraction of the angular momentum from the surface of the disk. Therefore, we introduce the following functional fit for the accretion torque as a function of $R_\text{t}/R_\star$:
\begin{align} \label{eq:acc_torque_2}
\begin{split}
	\dot{J}_\text{acc} = {}& - K_\text{acc} \lvert \dot{M}_\text{acc} \rvert (GM_\star R_\star)^{1/2} \left(\frac{R_\text{t}}{R_\star}\right)^{(1/2)+m_\text{acc}}, \\
\end{split}
\end{align}
where $K_\text{acc}$ and $m_\text{acc}$ are best-fit dimensionless parameters. 
In Figure~\ref{fig:Jdot_acc_with_Kacc_Rt_fitting_function}, we plot the accretion torque as a function of the right hand side of Equation~(\ref{eq:acc_torque_2}), where the best-fit dimensionless parameters $K_\text{acc}=0.775$ and $m_\text{acc}=-0.147$, giving a truncation radius power-law index of $(1/2)+m_\text{acc} = 0.353$ for our best-fit parameters. 
The accretion torque in our formulation predicts a lower value compared to the analytical formulation. For our parameter regime, one can predict the accretion torque using $\rho_{\text{d},\star}$, using the mass accretion rate parameterization in Appendix~\ref{sec:param_Mdot}.

\begin{figure}
\begin{center}
\includegraphics[width=0.4625\textwidth]{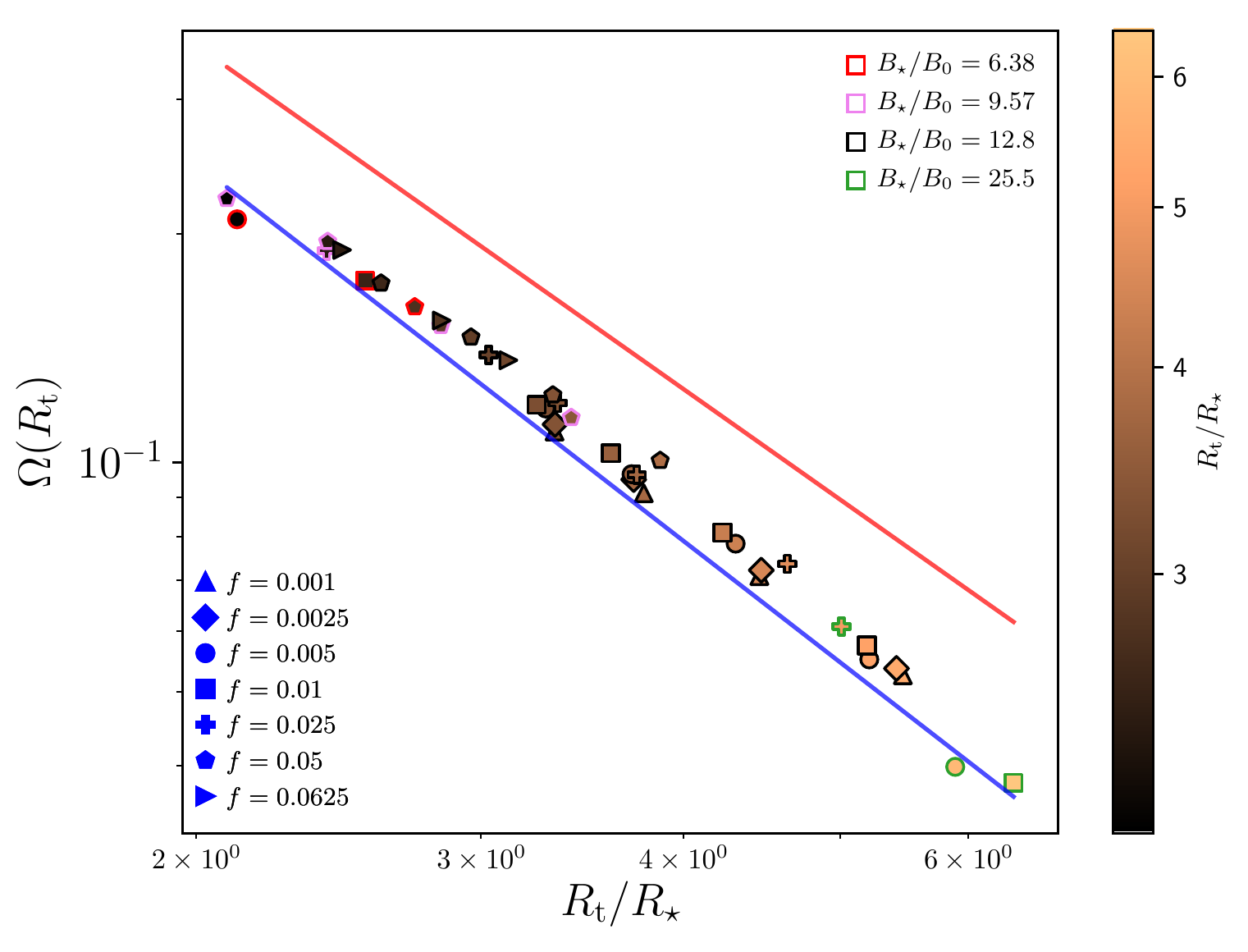}
\caption{Rotation rate of disk material at the truncation radius, $\Omega(R_\text{t})$, as a function of $R_\text{t}/R_\star$ for all simulations. The red line shows the Keplerian solution, indicating that all simulation cases exhibit sub-Keplerian rotation.  
The blue line shows the prediction of Equation~(\ref{eq:omega_Rt}), but using the fit coefficients for the accretion torque of Equation~(\ref{eq:acc_torque_2}).
The approximate match of the blue line to the data indicates that sub-Keplerian rotation is the primary reason that accretion torques in the SDI are less than the analytic predictions.
 \label{fig:Omega_Rt_v_Rt}}
\end{center}
\end{figure}

If the interpretation of \citet{refId0} is correct, the deviation of our formulation from the analytic one should be related to the disk rotation rate.  
Taken at face value, Equations (\ref{eq:acc_torque}) and (\ref{eq:acc_torque_2}) imply that the actual rotation of the disk at $R_\text{t}$ is
\begin{equation} \label{eq:omega_Rt}
  \begin{split}
	\Omega (R_\text{t}) &= 
	K_\text{acc} 
	\left({\frac{R_\text{t}}{R_\star}}\right)^{m_\text{acc}} \Omega_\text{K}(R_\text{t})  \\ 
	&= K_\text{acc} \left({\frac{GM_\star}{R_\star^3}}\right)^{1/2}  \left({\frac{R_\text{t}}{R_\star}}\right)^{-(3/2)+ m_\text{acc}}.
  \end{split}
\end{equation}
To test whether this is true, we computed $\Omega (R_\text{t})$ in our simulations using the time-averaged local value, and we plot these as a function of truncation radius in Figure~\ref{fig:Omega_Rt_v_Rt}.  
The red line shows the expectation for purely Keplerian rotation ($K_\text{acc} = 1$ and $m_\text{acc} = 0$), which indicates that the simulated disks rotate more slowly than Keplerian, and their rotation has a slightly different radial dependence.
The blue line shows the prediction of Equation~(\ref{eq:omega_Rt}), using the values of $K_\text{acc}$ and $m_\text{acc}$ that were fit to Equation~(\ref{eq:acc_torque_2}).  The approximate match between the blue line and data corroborates that the divergence of our accretion torque formulation from the analytical solution is primarily a result of the accretion disk rotating at sub-Keplerian velocity in the truncation region.
The match of the prediction to the simulation values is not perfect, which may be due to the difficulty to define a single value of $\Omega(R_\text{t})$ in the simulations or may suggest that other mechanisms contribute a small additional adjustment to the accretion torque. 

In a Keplerian disk, the corotation radius, $R_\text{co}$, is defined by $\Omega_\star = \Omega_\text{K}(R_\text{co})$.  For our simulations, we can now define the magnetic corotation radius as
\begin{equation} \label{eq:Rco_eff}
R_\text{co,m} = \left[K_\text{acc} \left(\frac{R_\text{t}}{R_\star}\right)^{m_\text{acc}}\right]^{2/3} R_\text{co}.
\end{equation}
It is clear that the location where the disk angular rotation rate equals that of the star is always smaller than the value for a Keplerian disk by a factor of $K_\text{acc} (R_\text{t}/R_\star)^{m_\text{acc}}$.  This value of $R_\text{co,m}$ is shown in Figures~\ref{fig:density_150} and \ref{fig:density_zoom}.

\subsection{Stellar magnetospheric ejection torque} \label{sec:ME_torque}

The MEs exert a torque on the disk and star, as well as carrying some angular momentum out of the system in an outflow.  
This is a complex and highly dynamic region.  
Our approach for finding a formulation for the torque on the star, due to MEs, is thus guided by the physics, but there are necessarily some heuristic aspects in the following formulations.
We start with an analysis of the ME torque on the surface of the disk.
The differential magnetic torque, due to the exertion of magnetic field lines threading an annulus of radial width $dR$ in a disk, is given as \citep[see, e.g.,][]{1996MNRAS.280..458A,2005MNRAS.356..167M}

\begin{equation} \label{eq:ME_torque_diff}
	\frac{d\dot{J}_\text{mag}}{dR} = q B_\theta^2 R^2,
\end{equation}
where $q$ {is a measure} of the twist of the field, expressed as the ratio of the toroidal to the vertical field at the disk surface, $B_\phi^+$ and $B_\theta$, respectively.
Next, we will assume that some fraction of the torque exerted by the MEs on the disk is also exerted on the star.
Recognizing that the size of the ME region ($\Delta R$) is small and bounded on one side by $R_\text{t}$, we can write the torque exerted on the star due to MEs as
\begin{equation} \label{eq:ME_torque}
	\dot{J}_{\text{ME},\star} \propto q(R_\text{t}) B_\theta(R_\text{t})^2 R_\text{t}^2 \Delta R.
\end{equation}
Thus, we need to include formulations for the magnetic twist ($q$), vertical field strength ($B_\theta$), and size of the ME region connected in the disk ($\Delta R$).

\begin{figure}
\begin{center}
\includegraphics[width=0.4625\textwidth]{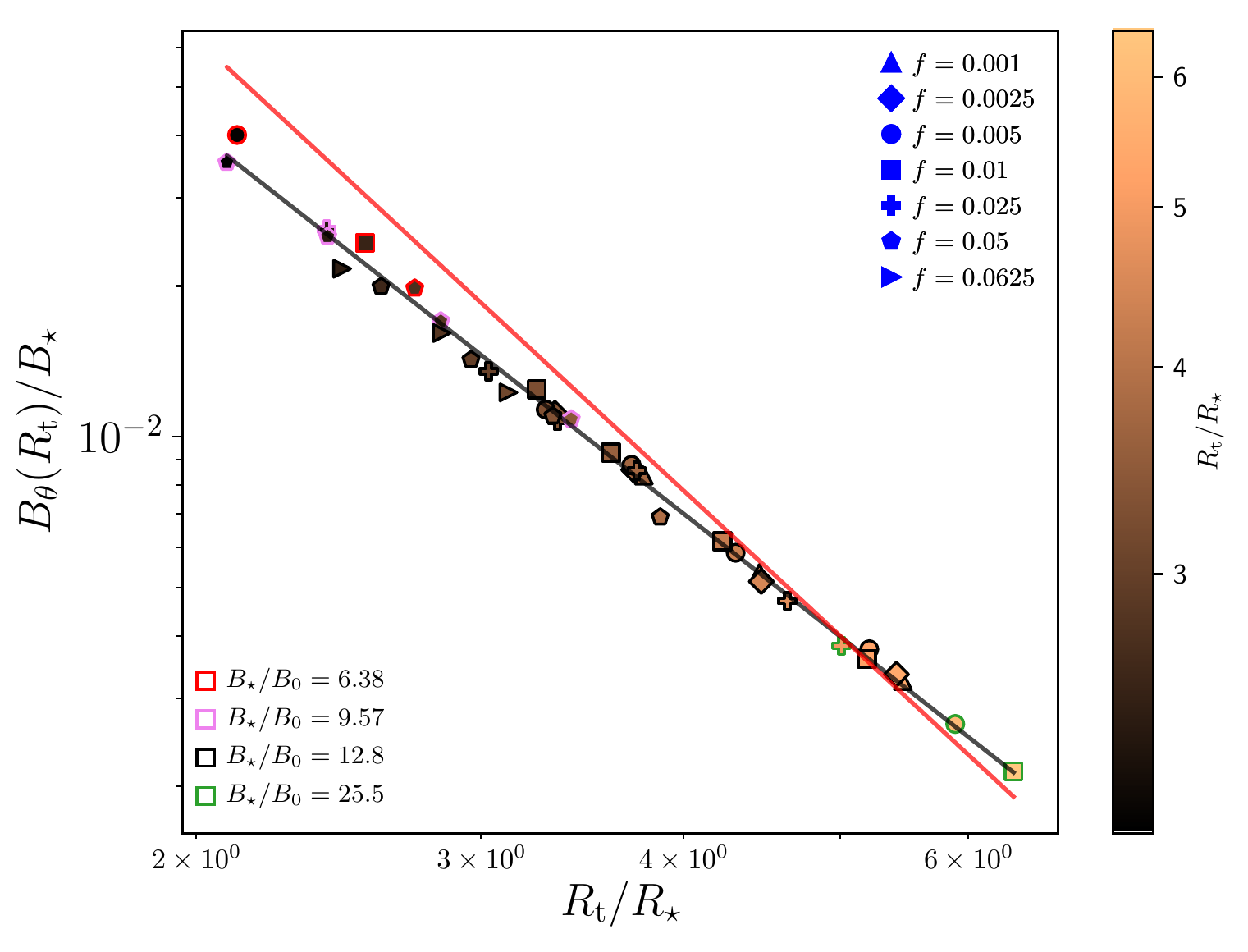}
\caption{Vertical magnetic field strength in the disk midplane,  $B_\theta (R_\text{t})/B_\star$, as a function of $R_\text{t}/R_\star$ for all simulations.  The red line shows a simple dipolar magnetic field ($B_\theta (R_\text{t})/B_\star = 0.5 (R_\text{t}/R_\star)^{-3}$), indicating the field perturbations in the SDI results in a shallower radial dependence.  The black line shows the fit to Equation~(\ref{eq:B_theta_Rt}), giving $B_\theta (R_\text{t})/B_\star = 0.236 (R_\text{t}/R_\star)^{-2.54}$.  \label{fig:Btheta_rt_v_Rt}}
\end{center}
\end{figure}

The field twist at the disk surface can be expressed in terms of the stellar and local rotation rates to account for the differential rotation between the MEs and the star \citep[see, e.g.,][]{1992MNRAS.259P..23L,1996MNRAS.280..458A,2005MNRAS.356..167M,2019A&A...632A...6G}:
\begin{equation} \label{eq:twist}
	q(R_\text{t}) \equiv - \left.{\frac{B_\phi^+}{B_\theta}}\right|_{R_\text{t}} \propto \left[\frac{\Omega_\star}{\Omega(R_\text{t})} - 1\right]
	\propto \left[\left(\frac{R_\text{t}}{R_\text{co,m}}\right)^{3/2} - 1\right].
\end{equation}
The polar magnetic field strength evaluated at the truncation radius can be written as
\begin{equation} \label{eq:B_theta_Rt}
	B_\theta(R_\text{t}) = K_B B_\star \left(\frac{R_\text{t}}{R_\star}\right)^{m_B},
\end{equation}
where $K_B$ and $m_B$ are best-fit dimensionless parameters. 
Figure~\ref{fig:Btheta_rt_v_Rt} shows $B_\theta (R_\text{t})/B_\star$ (computed using time-averages in our simulations) as a function of $R_\text{t}/R_\star$. The red line shows the expectation for a simple dipole ($K_B = 0.5, m_B=-3$, which also corresponds to our initial conditions), indicating that the SDI perturbs the field in the simulations to have a slightly different radial dependence.  
The black line shows the best fit, giving $K_B = 0.236$ and $m_B = -2.54$.
Finally, we assume the size of the ME connected region scales simply with $R_\text{t}/R_\star$: 
\begin{equation} \label{eq:DeltaR_Rt}
  \Delta R / R_\star \propto (R_\text{t}/R_\star)^{m_{\Delta R}},
\end{equation}
where the index $m_{\Delta R}$ can be inferred from our final formulation below.

\begin{figure}
\begin{center}
\includegraphics[width=0.4625\textwidth]{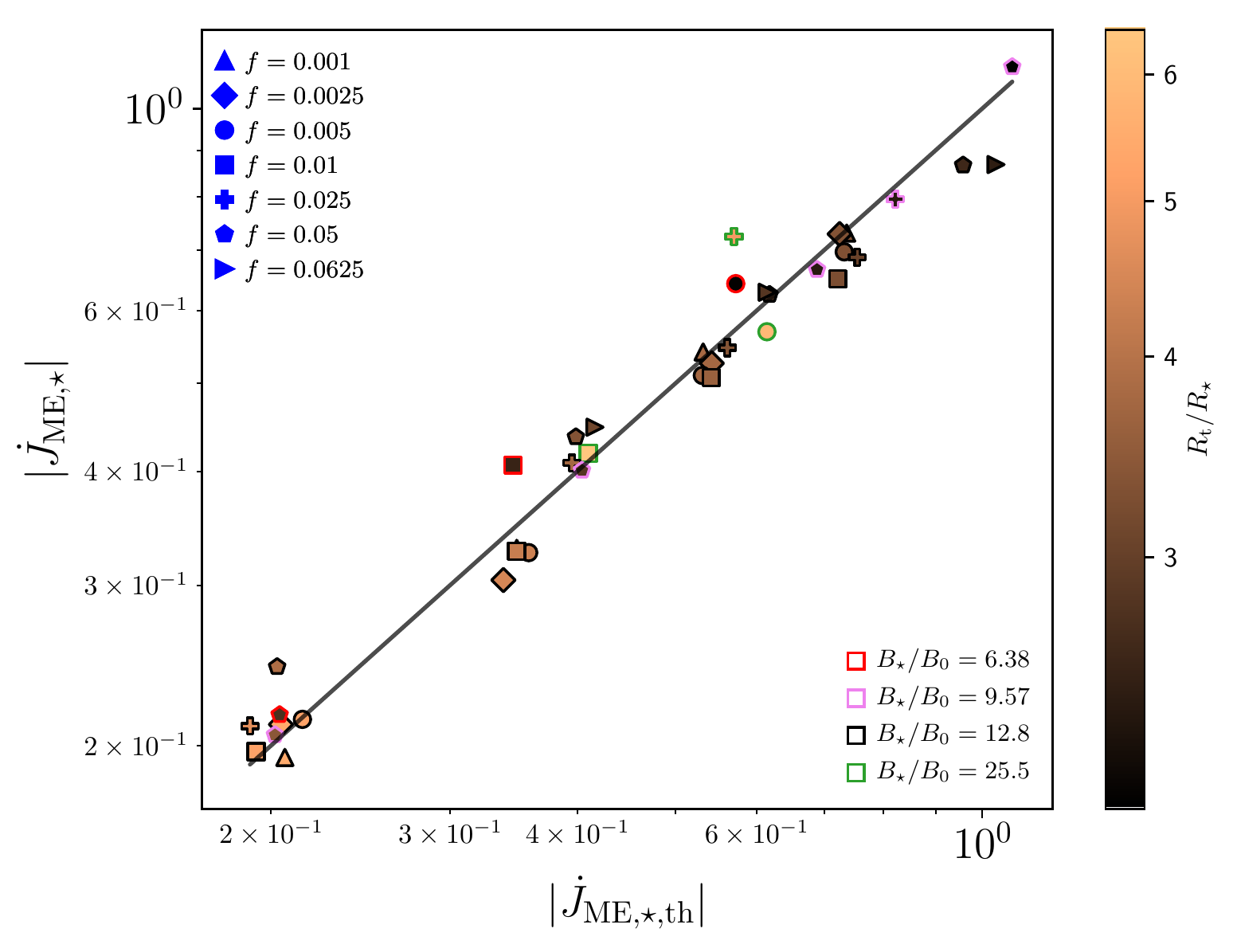}
\caption{The (absolute) stellar ME torque, $\lvert \dot{J}_{\text{ME},\star} \rvert$, as a function of the parameterization in Equation~(\ref{eq:ME_torque_2}) for all simulations. The black line shows $y=x$, illustrating the goodness of fit. \label{fig:Jdot_ME_KME_form}}
\end{center}
\end{figure}

Combining Equations~(\ref{eq:ME_torque})-(\ref{eq:DeltaR_Rt}), we express the stellar ME torque as
\begin{align} \label{eq:ME_torque_2}
\begin{split}
	\dot{J}_{\text{ME},\star} = {}& K_{\text{ME},\star} \left[\left(\frac{R_\text{t}}{R_\text{co,m}}\right)^{3/2} - 1\right] B_\star^2 R_\star^3 \left(\frac{R_\text{t}}{R_\star}\right)^{m_{\text{ME,}\star}}, \\
\end{split}
\end{align}
where $K_{\text{ME},\star}$ and $m_{\text{ME},\star}$ are best-fit dimensionless parameters.  
It is clear that $m_{\text{ME},\star}$ is equivalent to $2 + 2m_B+m_{\Delta R}$, and the constant $K_{\text{ME},\star}$ combines all of the proportionality constants above, as well as the fraction of the ME torque that is applied to the star\footnote{Note that the specific formulation/interpretation here is somewhat heuristic.  An alternative interpretation, for example, could assume a constant $\Delta R$ and instead that there is some $R_\text{t}$-dependence in the fraction of the ME torque applied to the star.}.
Figure~\ref{fig:Jdot_ME_KME_form} shows the stellar ME torque as a function of the right hand side of Equation~(\ref{eq:ME_torque_2}), where the best-fit dimensionless parameters are $K_{\text{ME},\star} = 0.00772$ and $m_{\text{ME},\star} = -2.54$. In all of our simulations, the MEs add angular momentum to the star. Combining the parameterization of the truncation radius (Equations~(\ref{eq:Rt_Y_acc})-(\ref{eq:Y_acc})), we can equivalently express the stellar ME torque in an alternative form, in terms of $R_\text{t}/R_\star$, $f$ and $\lvert \dot{M}_\text{acc} \rvert$:
\begin{align} \label{eq:ME_torque_2_Rt}
\begin{split}
	\dot{J}_{\text{ME},\star} = {}& \frac{4 \sqrt{2} \pi K_{\text{ME},\star}}{K_\text{t}^{1/m_\text{t}}} \lvert \dot{M}_\text{acc}\rvert (GM_\star R_\star)^{1/2}  \\ 
	& \left[\frac{f}{K_\text{acc}} \left(\frac{R_\text{t}}{R_\star}\right)^{(3/2)-m_\text{acc}} - 1\right] 
	\left(\frac{R_\text{t}}{R_\star}\right)^{m_{\text{ME},\star} + (1/m_\text{t})}, \\
\end{split}
\end{align}
giving a positive truncation radius power-law index of $m_{\text{ME},\star} + (1/m_\text{t}) = 0.401$ for our best-fit parameters. For our parameter regime, one can predict the stellar ME torque using $\rho_{\text{d},\star}$, using the mass accretion rate parameterization in Appendix~\ref{sec:param_Mdot}.

\subsection{Stellar wind torque} \label{sec:SW_torque}

The stellar wind torque is parameterized as
\begin{equation}\label{eq:torque_wind}
\dot{J}_\text{wind} = \dot{M}_\text{wind} \Omega_\star \langle r_\text{A} \rangle^2,
\end{equation}
where $\langle r_\text{A} \rangle$ is the torque-average Alfvén radius, which is actually defined by equation~(\ref{eq:torque_wind})  \citep[following][]{1993MNRAS.262..936W,2008ApJ...678.1109M}.
Thus, in our simulations, we compute $\langle r_\text{A} \rangle$ using
\begin{equation}\label{eq:RA_torque_wind}
\frac{\langle r_\text{A} \rangle}{R_\star} = \left(\frac{\dot{J}_\text{wind} }{ \dot{M}_\text{wind} \Omega_\star R_\star^2}\right)^{1/2}.
\end{equation}
This quantity represents an ``effective magnetic lever arm'' that quantifies the efficiency of the wind spin-down torque \citep[see, e.g.,][]{1967ApJ...148..217W}.
Values for $\langle r_\text{A} \rangle / R_\star$ in our simulations can be found in Table~\ref{tab:Pluto_sims}.

\begin{figure}
\begin{center}
\includegraphics[width=0.4625\textwidth]{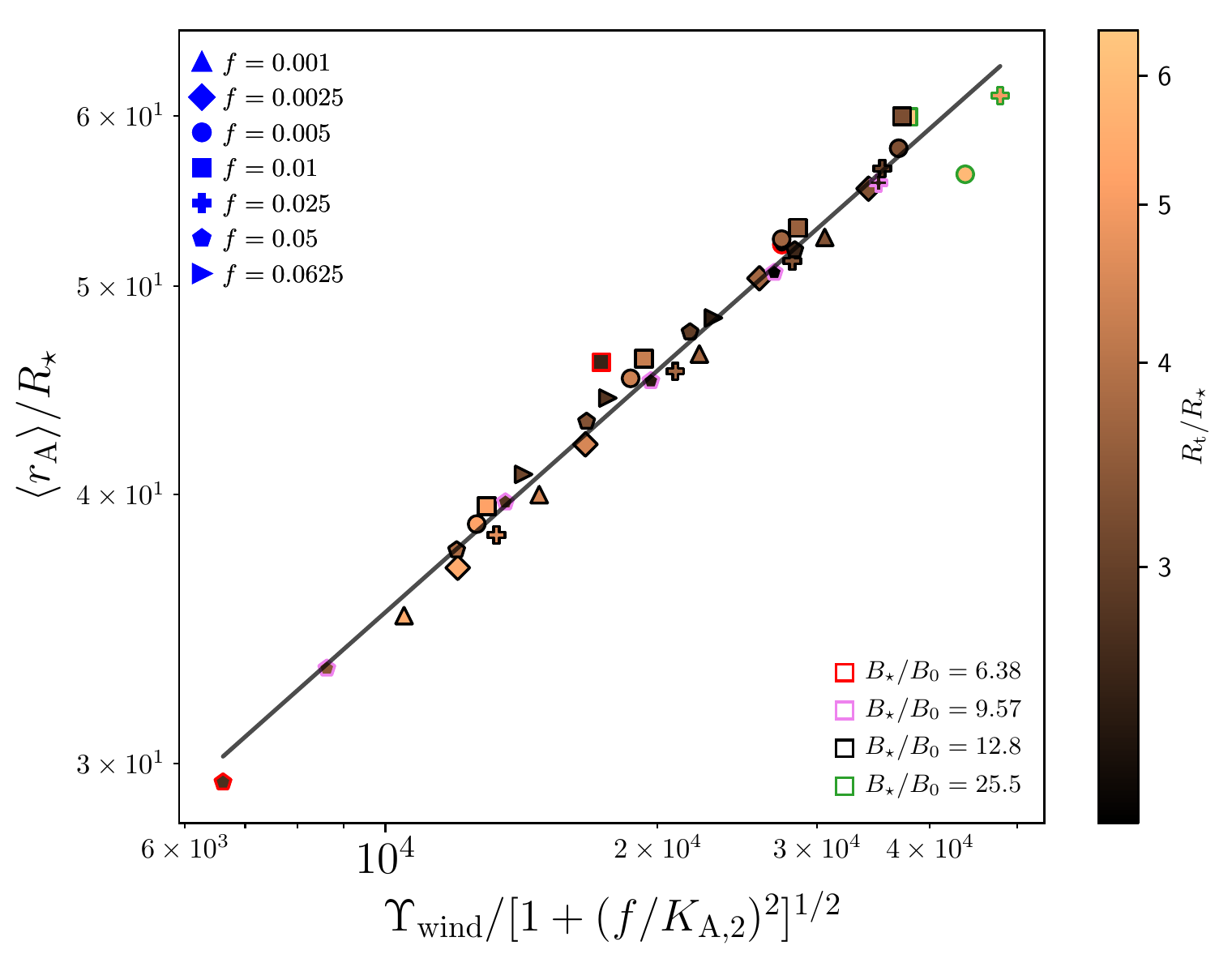}
\caption{Normalized Alfvén radius in the stellar wind, $\langle r_\text{A} \rangle/R_\star$, as a function of the wind open flux magnetization parameter, $\Upsilon_\text{wind} / [1 + (f/K_{\text{A},2})^2]^{1/2}$, for all simulations.  The black line shows the fit to Equation~(\ref{eq:Ra_Y_wind}), giving $\langle r_\text{A} \rangle/R_\star = 1.13 \{\Upsilon_\text{wind} / [1 + (f/0.0356)^2]^{1/2}\}^{0.373}$. \label{fig:R_A_v_Upsilon_wind}}
\end{center}
\end{figure}

A torque formulation for stellar wind simulations was developed by \citet{2015ApJ...798..116R}, which relates the Alfvén radius to the unsigned stellar wind (or open) magnetic flux. Using Equations~(\ref{eq:Br}) and~(\ref{eq:unsigned_flux}), the open flux, $\Phi_\text{wind}$, can be determined by integrating between opening angles enclosing the stellar wind region at the stellar surface, i.e., evaluating Equation~(\ref{eq:unsigned_flux}) at $R=R_\star$ between $0 \leq \theta \leq \theta_\text{wind,a}$ and $\theta_\text{wind,b} \leq \theta \leq \pi$, where $\theta_\text{wind,a}$ and $\theta_\text{wind,b}$ are the stellar wind anchoring angles for the northern and southern hemisphere, respectively. \citet{2015ApJ...798..116R} show that $\langle r_\text{A} \rangle / R_\star$ is related to $\Phi_\text{wind}$ via a single power-law, independent of magnetic geometry:
\begin{equation}\label{eq:Ra_Y_wind}
\frac{\langle r_\text{A}\rangle}{R_\star} = K_{\text{A},1} \left\{\frac{\Upsilon_\text{wind}}{[1 + (f/K_{\text{A},2})^2]^{1/2}}\right\}^{m_\text{A}},
\end{equation}
where $K_{\text{A},1}$, $K_{\text{A},2}$, and $m_\text{A}$ are best-fit dimensionless parameters, and
\begin{equation}\label{eq:Y_wind}
\Upsilon_\text{wind} = \frac{\Phi_\text{wind}^2}{4 \pi R_\star^2\dot{M}_\text{wind} v_\text{esc}}
\end{equation}
is the ``magnetization parameter'' of the stellar wind, based on the open magnetic flux. Values for $\Upsilon_\text{wind}$ in our simulations can be found in Table~\ref{tab:Pluto_sims}. 
Figure~\ref{fig:R_A_v_Upsilon_wind} shows $\langle r_\text{A} \rangle/R_\star$ as a function of $\Upsilon_\text{wind} / [1 + (f/K_{\text{A},2})^2]^{1/2}$, where the best-fit dimensionless parameters $K_{\text{A},1}=1.13$, $K_{\text{A},2}=0.0356$, and $m_\text{w}=0.373$. 
Increasing $\Phi_\text{wind}$ results in a increased Alfvén radius, giving a more efficient torque on the star (i.e., more torque for a given $\dot M_\text{wind} \Omega_\star$). 
The Alfvén radius is also inversely proportional to the stellar wind mass loss rate, which itself is dependent on the coronal temperature or density, and $\Phi_\text{wind}$.

Equation~(\ref{eq:Y_wind}) can also be written as
\begin{equation}\label{eq:Y_wind2}
\Upsilon_\text{wind} = \left(\frac{\Phi_\text{wind}}{\Phi_\star}\right)^2 \Upsilon_\star,
\end{equation}
where
\begin{equation}\label{eq:Y_old}
\Upsilon_\star = \frac{\Phi_\star^2}{4 \pi R_\star^2 \dot{M}_\text{wind} v_\text{esc}}.
\end{equation}
Therefore, in order to predict the stellar wind torque as a function of the properties at the stellar surface, given by $\Upsilon_\star$, we must have a formulation for the open magnetic flux, $\Phi_\text{wind}$.

In the SDI setup, the differential rotation between the star and disk opens magnetic surfaces at latitudes higher than where the field connects to $R_\text{t}$ \citep[e.g.][]{2002ApJ...565.1205U}.  
The flux opened in this way includes the stellar wind flux (region (1), Figure~\ref{fig:density_150}), $\Phi_\text{wind}$, and also the flux participating in MEs (region (2)), $\Phi_{\text{ME},\star}$, the latter of which opens and closes quasi-periodically.  
In our simulations, we find the sum of these fluxes to scale as
\begin{equation}\label{eq:wind_ME_flux_Upsilon_acc}
\frac{\Phi_\text{wind} + \Phi_{\text{ME},\star}}{\Phi_\star} = K_{\Phi} \left(\frac{R_\text{t}}{R_\star}\right)^{m_{\Phi}},
\end{equation}
where the best-fit dimensionless parameters are $K_{\Phi} = 1.26$ and $m_{\Phi} = -1.27$\footnote{If the field were a simple dipole, the analytic solution is $K_{\Phi}=1$ and $m_{\Phi}=-1$, so our best fit values reflect the perturbation of the magnetospheric geometry by the SDI.}. 

\begin{figure}
\begin{center}
\includegraphics[width=0.4625\textwidth]{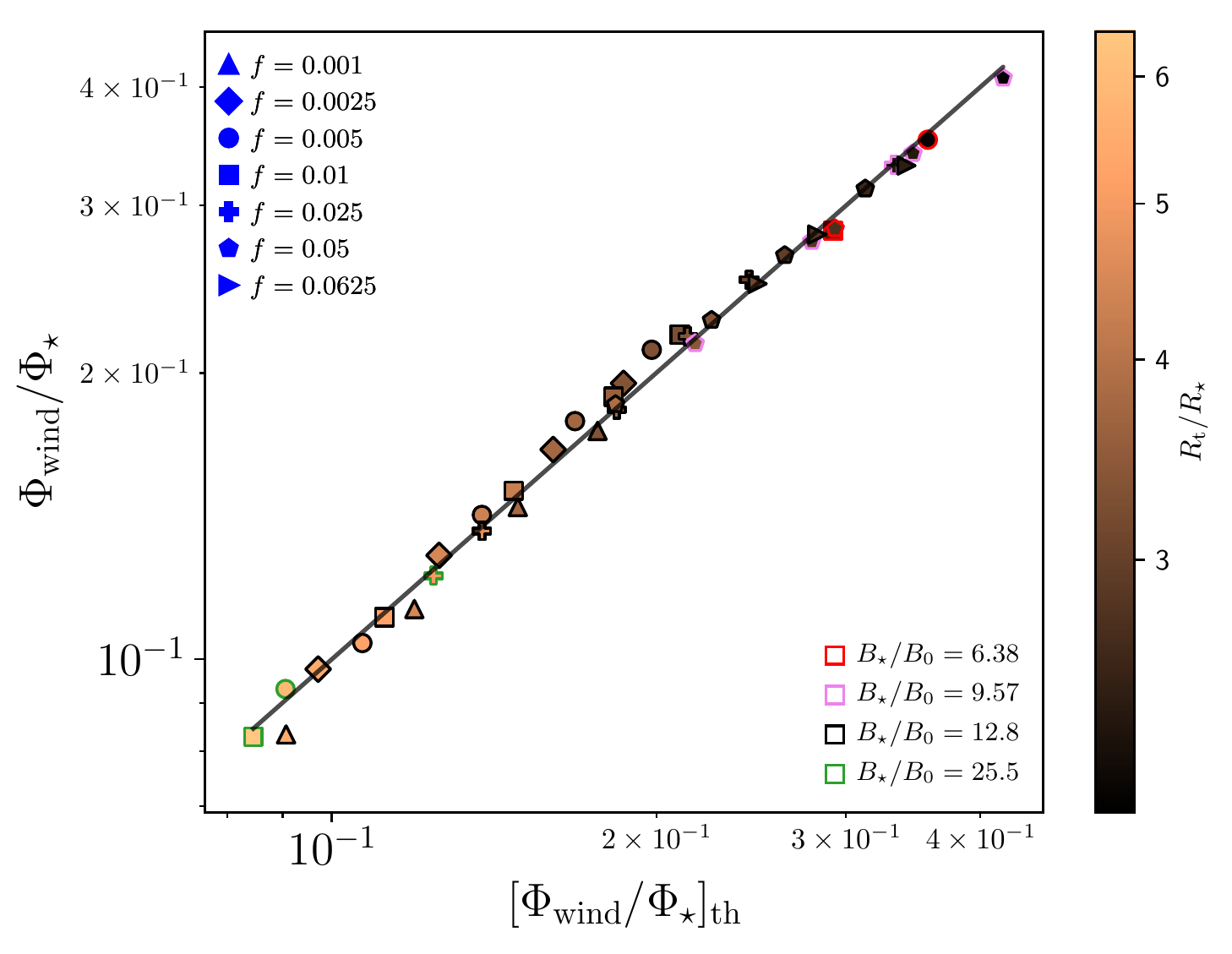}
\caption{Fraction of open flux in stellar wind, $\Phi_\text{wind}/\Phi_\star$, as a function of the parameterization in Equation~(\ref{eq:wind_flux_Upsilon_acc_f}) for all simulations. The black line shows $y=x$, indicating goodness of fit. \label{fig:phi_wind_div_star}}
\end{center}
\end{figure}

Due to the complexity of MEs, it is not obvious how to predict and remove their contribution in Equation~(\ref{eq:wind_ME_flux_Upsilon_acc}), although we expect $\Phi_{\text{ME},\star}$ depends on the magnetic twist and thus the spin rate of the star.  In practice, we tried several ideas and found the best agreement with the simple formulation
\begin{equation}\label{eq:wind_flux_Upsilon_acc_f}
\frac{\Phi_\text{wind}}{\Phi_\star} = K_{\Phi,1} \left(\frac{R_\text{t}}{R_\star}\right)^{m_{\Phi,1}} f^{m_{\Phi,2}}.
\end{equation}
Figure~\ref{fig:phi_wind_div_star} shows $\Phi_\text{wind}/\Phi_\star$ as a function of the right hand side of Equation~(\ref{eq:wind_flux_Upsilon_acc_f}), showing our simulations to follow this two-variable, power-law scaling with the best-fit parameters $K_{\Phi,1} = 1.36$, $m_{\Phi,1} = -1.34$ and $m_{\Phi,2} = 0.0614$.
Qualitatively, this suggests that, for a fixed $R_\text{t}$, the fraction of flux in the MEs increases for slower stellar rotation (smaller $f$), which also corresponds to more star-disk differential rotation in the truncation vicinity.
Therefore, for our SDI simulations, we find that $\Phi_\text{wind}/\Phi_\star$ strongly depends on the position of the truncation radius, and is independent of the stellar wind mass-loss rate, demonstrating that the accretion disk dominates over the wind dynamics.
The fact that the wind's open magnetic flux is independent of $\dot M_\text{wind}$ (and in fact depends on $\dot M_\text{acc}$, via $R_\text{t}$) represents a significant difference from the dynamics of ISWs (i.e., in systems without disks), and the implications of this are further discussed in Section~\ref{sec:discussion_wind}. 

\begin{figure}
\begin{center}
\includegraphics[width=0.4625\textwidth]{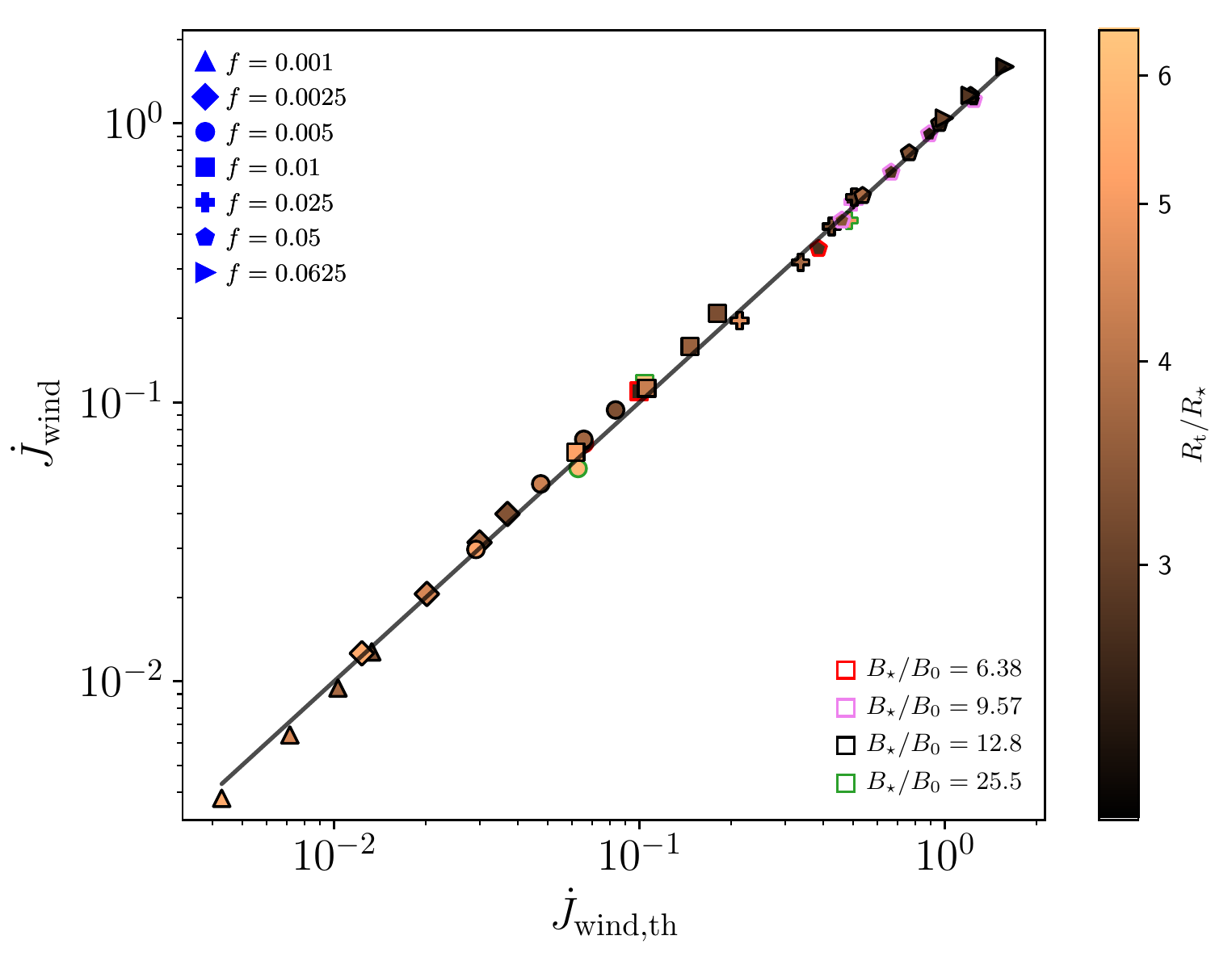}
\caption{The stellar wind torque, $\dot{J}_\text{wind}$, as a function of the parameterization in Equation~(\ref{eq:full_sw_torque}) for all simulations. The black line shows $y=x$, indicating goodness of fit. \label{fig:Jdot_wind}}
\end{center}
\end{figure}

Combining Equations~(\ref{eq:torque_wind}),~(\ref{eq:Ra_Y_wind}),~(\ref{eq:Y_wind2}), and~(\ref{eq:wind_flux_Upsilon_acc_f}), we express the stellar wind torque as
\begin{align}\label{eq:full_sw_torque_pre}
\begin{split}
\dot{J}_\text{wind} = {}& K_{\text{A},1}^2 K_{\Phi,1}^{4m_\text{A}} \dot{M}_\text{wind} (GMR_\star)^{1/2} f^{1+4 m_\text{A} m_{\Phi,2}} \left(\frac{R_\text{t}}{R_\star}\right)^{4m_\text{A} m_{\Phi,1}} \\
& \left\{\frac{\Upsilon_\star}{[1 + (f/K_{\text{A},2})^2]^{1/2}}\right\}^{2m_\text{A}}
\end{split}
\end{align}
Using Equations~(\ref{eq:unsigned_flux}),~(\ref{eq:Rt_Y_acc}), and~(\ref{eq:Y_acc}), we can equivalently express the stellar wind torque (Equation~(\ref{eq:full_sw_torque_pre})) in terms of $R_\text{t}/R_\star$ and the mass ejection efficiency $\dot{M}_\text{wind}/\lvert\dot{M}_\text{acc}\rvert$:
\begin{align}\label{eq:full_sw_torque}
\begin{split}
\dot{J}_\text{wind} = {}& \frac{ K_{\text{A},1}^2 (\alpha \pi K_{\Phi,1})^{4m_\text{A}} }{K_\text{t}^{2m_\text{A}/m_\text{t}}} \lvert \dot{M}_\text{acc} \rvert (GM_\star R_\star)^{1/2} \left(\frac{\dot{M}_\text{wind}}{\lvert \dot{M}_\text{acc} \rvert}\right)^{1-2m_\text{A}} \\
& f^{1+4 m_\text{A} m_{\Phi,2}} \left[1 + \left(\frac{f}{K_{\text{A},2}}\right)^2\right]^{-m_\text{A}} \left(\frac{R_\text{t}}{R_\star}\right)^{4 m_\text{A} m_{\Phi,1} + (2m_\text{A}/m_\text{t})},
\end{split}
\end{align}
where $\alpha=2$ for our dipolar configuration, giving a positive truncation radius power-law index of $4 m_\text{A} m_{\Phi,1} + (2m_\text{A}/m_\text{t}) = 0.194$ for our best-fit parameters. 
Figure~\ref{fig:Jdot_wind} shows the stellar wind torque as a function of the right hand side of Equation~(\ref{eq:full_sw_torque}), showing excellent correspondence with our stellar wind torque formulation. 

\begin{deluxetable}{cccc}

\tablecaption{Best-fit dimensionless parameters from scaling laws in Section~\ref{sec:torque_formulation}.\label{tab:dimensionless_params}}

\tablehead{Formulation & Parameter & Value & Equation}

\startdata
$R_\text{t}/R_\star$ & $K_\text{t}$ & 0.756 & (\ref{eq:Rt_Y_acc}) \\
 & $m_\text{t}$ & 0.340 &  \\
\hline
$\dot{J}_\text{acc}$ & $K_\text{acc}$ & 0.775 & (\ref{eq:acc_torque_2}) \\
 & $m_\text{acc}$ & -0.147 &  \\
\hline
$B_\theta(R_\text{t})$ & $K_B$ & 0.236 & (\ref{eq:B_theta_Rt}) \\
 & $m_B$ & -2.54 &  \\
\hline
$\dot{J}_{\text{ME},\star}$ & $K_{\text{ME},\star}$ & 0.00772 & (\ref{eq:ME_torque_2}) \\
 & $m_{\text{ME},\star}$ & -2.54 &  \\
\hline
$\langle r_\text{A} \rangle /R_\star$ & $K_\text{A,1}$ & 1.13 & (\ref{eq:Ra_Y_wind}) \\
 & $K_\text{A,2}$ & 0.0356 & \\
 & $m_\text{A}$ & 0.373 & \\
\hline
$\Phi_\text{wind}/\Phi_\star$ & $K_{\Phi,1}$ & 1.36 & (\ref{eq:wind_flux_Upsilon_acc_f}) \\
 & $m_{\Phi,1}$ & -1.34 & \\
 & $m_{\Phi,2}$ & 0.0614 & \\
\hline
$\dot{M}_\text{acc}$ & $K_M$ & 0.00221 & (\ref{eq:Mdot_acc_param}) \\
\enddata
\end{deluxetable}

\section{Discussion and conclusions}\label{sec:dis_conc}

In this work, we perform 38 2.5D axisymmetric MHD SDI simulations to investigate how the net torque experienced by the star is impacted by variations in the stellar rotation rate (expressed as the stellar break-up fraction $f$), the stellar surface polar magnetic field strength ($B_\star$), and the mass accretion rate (by changing the initial accretion disk density ($\rho_{\text{d,}\star}$)). 
All of our simulations are in a net spin-up regime. {In this parameter study, we are motivated to further investigate the transport of angular momentum to and from the star via MEs during its PMS phase}, and to produce improved torque formulations for the three mechanisms influencing the net stellar torque, namely, the stellar wind, the accretion, and the MEs. We create torque formulations for each, allowing for the investigation of spin-evolution of PMS objects using 1D stellar evolution codes. In Appendix~\ref{sec:stellar_evo_code}, we write a guide for implementing our torque formulations in a stellar evolution code, and discuss the range of validity when using our simulations.

\subsection{Truncation radius}\label{sec:discussion_Rt}

We express a parameterization of the truncation radius with the ``disk magnetization" parameter, $\Upsilon_\text{acc}$, where the proportionality constant $K_\text{t} = 0.756$ and the power-law index $m_\text{t} = 0.340$, which scales with $B_\star$ and inversely with the mass accretion rate. This is a steeper scaling compared to the analytical power-law index ($m_\text{t} = 2/7$), due to the fact that the accretion disk perturbs the magnetosphere. 
Using an identical set up, \citet{2020arXiv200900940P} find $K_\text{t}=0.704$ (corrected for polar magnetic field strength) and $m_\text{t}=0.350$ for their $B_\star$ parameter study, showing close agreement with our determined best-fit dimensionless parameters. \citet{2019A&A...632A...6G} assume a dipolar magnetosphere and use $K_\text{t} = 0.766$ and $m_\text{t} = 2/7$ for their $R_\text{t}$ parameterization (corrected for polar magnetic field strength), thus, we predict higher truncation radii than those predicted by \citet{2019A&A...632A...6G} for our parameter range. \citet{2013MNRAS.433.3048K} found a much smaller $m_\text{t}=0.2$ in their 3D simulations of magnetospheric accretion, for models that have a dipolar magnetic field tilted $30^\circ$ about the rotational axis. Therefore, further investigation into magnetic field inclination and its influence on $m_\text{t}$ would be required to make a direct comparison.

In Appendix~\ref{sec:param_Mdot}, we parameterize the mass accretion rate in terms of our three input parameters, allowing for the prediction of the truncation radius, accretion torque, and stellar ME torque, as a function of $\rho_{\text{d},\star}$, rather than $\dot{M}_\text{acc}$.

\subsection{Accretion torque}\label{sec:discussion_acc}

For the accretion torque, we propose a formulation that considers the sub-Keplerianity of the accretion disk, due to the extraction of a fraction of angular momentum from the disk caused by MEs. We use a sub-Keplerian correction term $K_\text{acc}(R_\text{t}/R_\star)^{m_\text{acc}}(<1)$ to represent the accretion torque in terms of the fraction of the analytical accretion torque (which assumes the accretion disk is rotating at Keplerian velocity), i.e., $\dot{J}_\text{acc} = K_\text{acc}(R_\text{t}/R_\star)^{m_\text{acc}} \dot{J}_\text{acc,K}$. From this, we define a magnetic corotation radius $R_\text{co,m}<R_\text{co}$, where the magnetic twist becomes zero at $R_\text{t}=R_\text{co,m}$. 
For our simulations, $0.590 \leq K_\text{acc}(R_\text{t}/R_\star)^{m_\text{acc}} \leq 0.695$, which is higher than the sub-Keplerian coefficient of $K_\text{acc}\approx 0.4$, with no additional dependence on $R_\text{t}$, that was used in the spin-evolution calculations of \citet{2019A&A...632A...6G}; we comment on the choice of coefficients in \citet{2019A&A...632A...6G} in Section~\ref{sec:compar_lit}.
Therefore, we predict a higher spin-up accretion torque for a given SDI case in our parameter regime. \citet{2020arXiv200900940P} adopted a fixed $K_\text{acc} = 0.790$ with no additional dependence on $R_\text{t}$, predicting an even higher accretion torque than our formulation; however, this value was determined using a smaller set of simulations, solely changing $B_\star$.

\subsection{Stellar ME torque}\label{sec:discussion_ME}

Our stellar ME torque formulation offers additional self-consistency to the prescription introduced in \citet{2019A&A...632A...6G}, and adopted by \citet{2020arXiv200900940P}. 
Firstly, \citet{2019A&A...632A...6G} use the term $K_\text{rot}$ to account for the sub-Keplerian correction to the magnetic field twist (or the star-disk differential rotation). However, our $K_\text{acc} (R_\text{t}/R_\star)^{m_\text{acc}}$ parameter can be used interchangeably to represent both the accretion torque and field twist corrections. Their value for $K_\text{rot}=0.7$ is at the upper limit of our $K_\text{acc} (R_\text{t}/R_\star)^{m_\text{acc}}$ range, corresponding to our low $R_\text{t}/R_\star \sim 2$ cases. \citet{2019A&A...632A...6G} also assume a dipolar magnetic field when expressing the polar magnetic field strength at the truncation radius, whereas we employ the determined radial dependence (at the truncation radius) of the perturbed magnetic field structure. Finally, \citet{2019A&A...632A...6G} assume a linear dependence between the radial width $\Delta R$ and the truncation radius, whereas we allow the dependence on the truncation radius to be fitted. Their formulation also assumes a relatively higher $K_{\text{ME,}\star} = 0.0525$ (corrected for dipolar magnetic field strength) compared to our best-fit $K_{\text{ME,}\star} = 0.0072$. Overall, we predict a lower spin-up stellar ME torque for a given SDI case in our parameter regime.

All simulations demonstrate a spin-up stellar ME torque, with our truncation radii well within their respective magnetic corotation radii. 
We observe the largest scatter between simulation and derived values of the stellar ME torque compared to other torque formulations, which is likely due to the periodicity of the stellar ME torque affecting our time-averaged value, or from added complexity not considered in our stellar ME torque parameterization.

\subsection{Stellar wind torque}\label{sec:discussion_wind}

We determine a stellar wind torque formulation through the parameterization of the effective Alfvén radius, based on the open flux formulation employed by \citet{2015ApJ...798..116R}.  
For a known amount of open flux, our simulations follow a $\langle r_\text{A} \rangle$-$\Upsilon_\text{wind}$-$f$ relationship (Equation~(\ref{eq:Ra_Y_wind})) with a power-law index of $m_\text{A} = 0.373$.
ISW simulations with an initial dipolar configuration have been shown to produce power-law indices of $0.310 \leq m_\text{A} \leq 0.371$ \citep[see, e.g.,][]{2015ApJ...798..116R,2017ApJ...845...46F,2017ApJ...849...83P}. The power-law index depends on the wind acceleration, which can vary from the wind channel being modified by the presence of MEs, but also vary between numerical studies, e.g., due to differences in coronal temperature. A direct comparison of ISW and SDI simulations in \citet{2020arXiv200900940P}, exploring the $B_\star$ parameter space, produces power-law indices of $m_\text{A} = 0.355$ and $m_\text{A} = 0.439$, respectively. Qualitatively, this is a result of the accretion disk modifying the stellar wind geometry and acceleration; a more detailed comparison between SDI and ISW systems can be found in \citet{2020arXiv200900940P}. The deviations in their $m_\text{A}$ value from ours is likely within the systematic uncertainties of these simulations. Generally, we expect our values to be more precise because they are based on a higher number of simulations and varied parameters. In addition, the rotational-dependence in Equation~(\ref{eq:Ra_Y_wind}) introduces added complexity, that may slightly modify the power-law index, compared to the formulation of \citet{2020arXiv200900940P}.

{In order to compute the stellar wind torques in practice, one must be able to predict the open flux based on the magnetic properties at the stellar surface (e.g., $B_\star$) and other system properties.
In ISW simulations, the amount of open flux depends on the wind mass loss rate; specifically, $\Phi_\text{wind}/\Phi_\star$ inversely scales with $\Upsilon_\star$ \citep[see, e.g.,][]{2015ApJ...798..116R,2017ApJ...845...46F,2018ApJ...854...78F,2017ApJ...849...83P}.
By contrast, our simulations show that, in the presence of an accretion disk (and in the regime of $\dot{M}_\text{wind}/\lvert \dot{M}_\text{acc}\rvert \lesssim 1\%$), the open flux depends primarily on $R_\text{t}$ (and thus $\dot M_\text{acc})$ and results in an increase of the open flux, compared to an ISW case.
The apparent scaling of the open flux with $\Upsilon_\star$ in the SDI simulations of \citet{2020arXiv200900940P} is a consequence of a relationship between the mass accretion rate and wind mass-loss rate, due to the fact that they only vary $B_\star$; solely increasing $B_\star$ causes both mass fluxes to increase.
The independence of the open flux on $\dot M_\text{wind}$ (and its dependence instead on $\dot M_\text{acc}$) represents a significant difference from the dynamics of ISWs.
Once the amount of open flux is determined, the stellar wind flows along all flux that is available, extracting angular momentum.
The somewhat surprising result of this is that the stellar wind torque in SDI systems depends upon the accretion rate onto the star, in addition to the usual stellar wind parameters.  
As shown in Section~\ref{sec:SW_torque}, for SDI cases, $\dot{J}_\text{wind} \propto f^{1.09} \dot{M}_\text{wind}^{0.254} \lvert \dot{M}_\text{acc} \rvert^{0.680} B_\star^{0.132}$ (using best-fit parameters from Table~\ref{tab:dimensionless_params} and neglecting the wind centrifugal term\footnote{The wind centrifugal term is $[1+(f/K_\text{A,2})^2]^{-m_\text{A}}$, and is generally equal to nearly unity.}), whereas for ISW cases, ${\dot{J}_\text{wind}}\rvert_\text{ISW} \propto f \dot{M}_\text{wind}^{0.56} B_\star^{0.88}$ \citep[e.g.,][]{2012ApJ...754L..26M}.
For fixed stellar wind parameters, the stellar wind torque in the SDI increases when $\dot{M}_\text{acc}$ increases, due to a decrease in $R_\text{t}$ and thus increase in open magnetic flux.
The dependence of the wind torque on magnetic field strength is also much weaker than for ISWs because, in SDI systems, the fractional open flux decreases almost linearly with an increase in $B_\star$ (due to the sensitivity of fractional open flux on $R_\text{t}$ and $R_\text{t}$ on $B_\star$), so that the total amount of open flux is nearly independent of $B_\star$.}

{Since our simulations exhibit more open magnetic flux than ISW cases with the same wind parameters, their stellar wind torques are more efficient (i.e., they have more torque for a given $B_\star$ and $\dot M_\text{wind}$) than ISWs.  
{This could not be the case for high mass loss rates and/or large truncation radii. For example, comparing our SDI formulation to the ISW formulation in \citet{2017ApJ...845...46F} with $f=0.05$, for $R_\text{t}/R_\star \gtrsim 3$ and $\dot{M}_\text{wind}/\lvert \dot{M}_\text{acc}\rvert \gtrsim 10\%$, the torque of a stellar wind in an SDI system becomes less efficient than an ISW with the same mass loss rate and magnetic field. This is mainly due to the fact that in these conditions, according to the derived scalings, a more massive stellar wind could open more magnetic flux than the SDI.}
However, our formulations are based on simulations with $\dot{M}_\text{wind}/\lvert \dot{M}_\text{acc}\rvert \lesssim 1\%$, so it is not clear if our formulations can be extrapolated to $\sim 10\%$.
It is possible that when $\dot{M}_\text{wind}/\lvert \dot{M}_\text{acc}\rvert$ is large enough, the stellar wind could again contribute to the opening of the magnetic flux also in SDI systems, but when this transition occurs remains unknown.
Thus, until the regime of high $\dot{M}_\text{wind}/\lvert \dot{M}_\text{acc}\rvert$ is explored with simulations, caution should be used when extrapolating our formulations.}
{For clarity, we note that we are solely interested in how stellar wind torques scale in the presence of SDI; comparing to the ISW case gives us an indication of potential limits to our SDI scalings, when the wind becomes ``important enough" relative to the disk (so that the stellar wind torque starts to scale like an ISW).} 

{For the stellar wind torque, \citet{2019A&A...632A...6G} uses the \citet{2012ApJ...754L..26M} ISW prescription with a $30 \%$ higher value for $K_1$.
However, their formulation is not sensitive to the accretion rate, resulting in increasing underestimations of the stellar wind torque as $\dot M_\text{acc}$ increases. Therefore, we predict a higher spin-down stellar wind torque for a given SDI case in our parameter regime.} 

\subsection{Spin-equilibrium solution} \label{sec:compar_lit}

\begin{figure}
\begin{center}
\includegraphics[width=0.4625\textwidth]{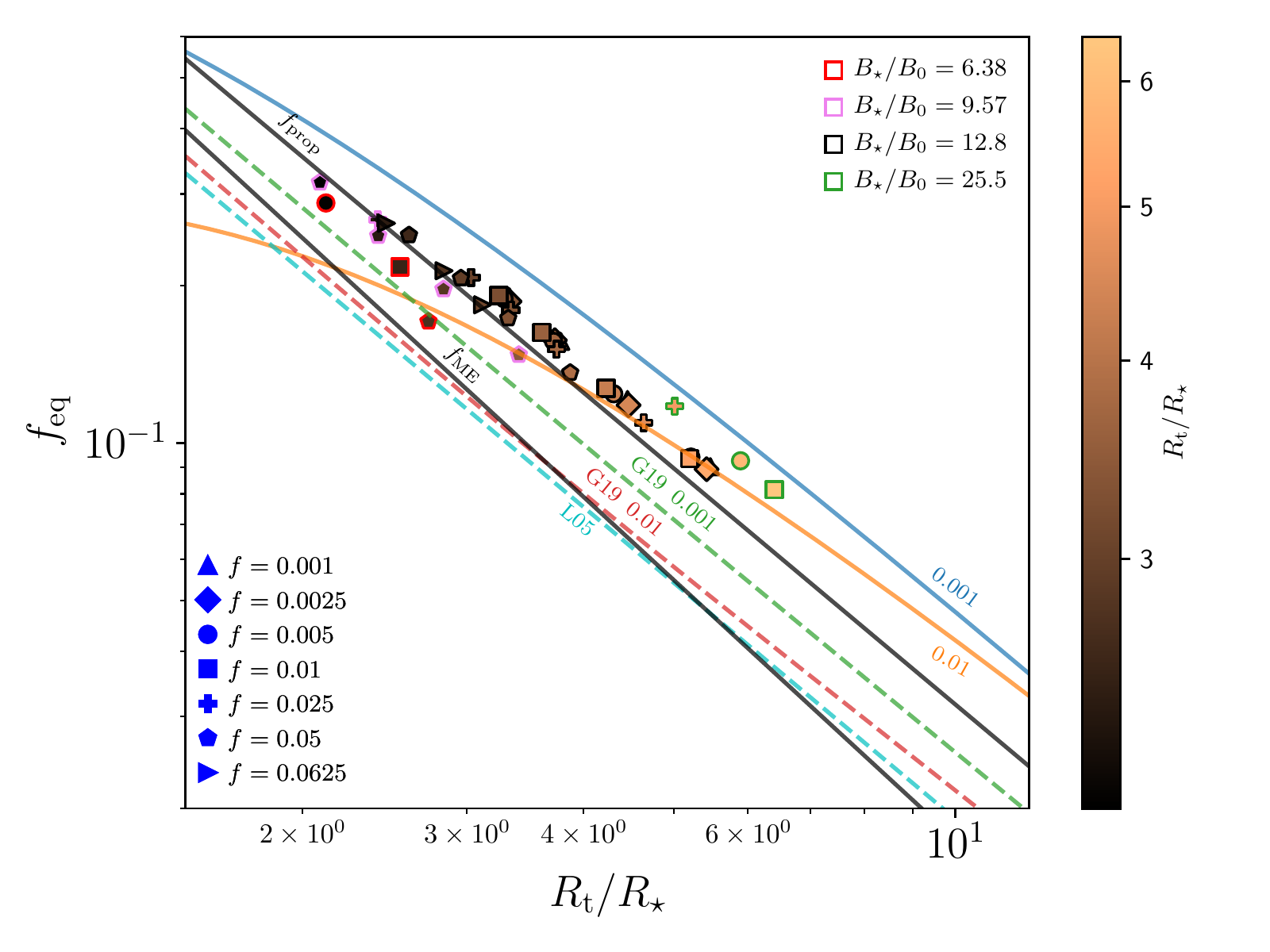}
\caption{$f_\text{eq}$ as a function of $R_\text{t}/R_\star$ for all simulations, and for $\dot{M}_\text{wind}/\lvert\dot{M}_\text{acc}\rvert = 0.001$ (blue line) and $0.01$ (orange line), using Appendix~\ref{sec:spin_eq}. We plot $f_\text{ME}$ (black solid line) to differentiate between spin-up and spin-down stellar ME torque regions, and $f_\text{prop}$ (black solid line) to indicate the transition into the propeller regime. Comparisons are made with spin-equilibrium solutions from \citet{2005ApJ...634.1214L} (cyan dashed line), and \citet{2019A&A...632A...6G} for $\dot{M}_\text{wind}/\lvert\dot{M}_\text{acc}\rvert = 0.001$ (green dashed line) and $0.01$ (red dashed line). \label{fig:f_eq_v_R_t}}
\end{center}
\end{figure}

We use our torque formulation to explore outside the parameter regime studied and predict equilibrium conditions for our simulations, defined as the conditions where the net stellar torque is zero, i.e., $\dot{J}_\text{wind} + \dot{J}_\text{acc} + \dot{J}_{\text{ME},\star} = 0$. 
We give the formulation for the equilibrium state in Appendix~\ref{sec:spin_eq}, where we show that the equilibrium stellar spin rate (as a fraction of break-up), $f_\text{eq}$, can be written in terms of the truncation radius and the mass ejection efficiency alone. 
In Figure~\ref{fig:f_eq_v_R_t}, we plot $f_\text{eq}$ as a function of $R_\text{t}/R_\star$.
The symbols show our simulation values, but plotted at a spin rate the star would need to have in order for the net torque in each simulation to be zero.
The blue and orange solid lines show continuous solutions for $\dot{M}_\text{wind}/\lvert\dot{M}_\text{acc}\rvert = 0.001$ and $0.01$ (using Equation~(\ref{eq:spin_eq_sol})), encapsulating a majority of our simulations. The apparent approximate inverse relationship with truncation radius illustrates that spin-equilibrium can be reached at lower spin rates as the effective magnetization (relative to the mass accretion rate) increases. 

The lower black line shows $f_\text{ME}=K_\text{acc} (R_\text{t}/R_\star)^{-(3/2)+m_\text{acc}}$, which illustrates the transition from a spin-up (below line) and spin-down (above line) stellar ME torque contribution for these spin-equilibrium cases. This $\dot{J}_{\text{ME},\star} = 0$ condition corresponds to a situation in which the disk is still sub-Keplerian so that, in principle, it can still steadily accrete. We also plot $f_\text{prop}=(R_\text{t}/R_\star)^{-3/2}$, which illustrates the transition into the propeller regime (above line), which corresponds to a situation in which the stellar centrifugal barrier hinders accretion by triggering super-Keplerian rotation.

Our formulation predicts that, in order for our simulated cases to have a net zero torque, a majority of our simulations must be in the propeller regime.
In these cases, the stellar ME torque (Equation~(\ref{eq:ME_torque_2})) in fact provides a spin-down torque onto the star, predicting a slower spin rate than if MEs were neglected. 
{In Figure~\ref{fig:f_eq_v_R_t}, the $\dot{M}_\text{wind}/\lvert\dot{M}_\text{acc}\rvert = 0.01$ solution (orange line) sits below the propeller regime transition (black line) when $R_\text{t}/R_\star < 4$. Therefore, our spin-equilibrium solutions suggest that it is possible to be in a spin-down stellar ME torque configuration without being in the propeller regime.} Since our spin-equilibrium solutions either exist in the propeller regime or have a spin-down stellar ME torque configuration, it is not possible to actually verify whether our formulation is valid in such a regime at present. 
It is possible that the open flux could saturate for $R_\text{t} \gtrsim R_{\text{co,K}}$, since it is unlikely to have steady accretion beyond this limit; even if accretion becomes intermittent, it is necessary that $R_\text{t} < R_{\text{co,K}}$, even just temporarily.

{The $J_\star = 0$ solution using the torque formulation in \citet{2019A&A...632A...6G} (labeled G19) predicts slower values of $f_\text{eq}$ compared to our simulations (except for very small $R_\text{t}$). {Overall, our formulation predicts a lower total spin-up torque (accretion plus MEs) and a higher total spin-down torque (stellar wind), giving a lower net spin-up stellar torque compared to the formulation in \citet{2019A&A...632A...6G}.} Therefore, for $\dot{M}_\text{wind}/\lvert\dot{M}_\text{acc}\rvert = 0.01$, the lower $f_\text{eq}$ values provided by \citet{2019A&A...632A...6G} are mostly due to the lower accretion torque. For $\dot{M}_\text{wind}/\lvert\dot{M}_\text{acc}\rvert = 0.001$, the more efficient spin-down torque from the MEs in the \citet{2019A&A...632A...6G} formulation contributes significantly to decrease the $f_\text{eq}$ value.
The solution in \citet{2005ApJ...634.1214L} (labeled L05), where $R_\text{co} = 1.4 R_\text{t}$, predicts slower values of $f_\text{eq}$ compared to our simulations. This also demonstrates that the \citet{2019A&A...632A...6G} solution is consistent with the $R_\text{t} \propto R_\text{co}$ scaling, proposed by this work.}

As an alternative to computing $f_\text{eq}$, we can also fix $f$ and $R_\text{t}$ at the value of each simulation and then solve for which value of mass ejection efficiency ($\dot{M}_\text{wind}/\lvert \dot{M}_\text{acc}\rvert$) gives a net zero torque.
Doing this, we find our formulation to suggest that $\dot{M}_\text{wind}/\lvert \dot{M}_\text{acc}\rvert \gtrsim 30 \%$ is required to achieve spin-equilibrium in our simulated cases, with the lowest mass-fraction corresponding to our most rapidly rotating cases ($f=0.0625$).
Using $R_\star$, $M_\star$, $B_\star$, $P_\star$, and $\dot{M}_\text{acc}$ observations of BP Tau as a representative CTTS case \citep[see][]{2014MNRAS.437.3202J,2019A&A...632A...6G}, our formulation suggests $\dot{M}_\text{wind}/\lvert \dot{M}_\text{acc}\rvert$ would need to be $\approx 25 \%$ for this star to be in spin-equilibrium. {We are {skeptical} of such high mass loss rates in CTTS; assuming that these high mass loss rates are not possible, the implication of our current torque formulation is that BP Tau could in fact be spinning up. However, this result is a prediction obtained solely via extrapolation of our current torque formulation, which may not be compatible within other parameter regimes, e.g., for stars in the propeller regime. The validity of our formulation outside of the low mass loss regime must be investigated in future work.}

\subsection{Limitations}\label{sec:limit}

Due to the complexity of SDI systems, various limitations of our study are listed hereafter. Firstly, the accretion disk, and its ability to transfer angular momentum, is treated using simple viscosity and diffusion formulations, with constant anomalous coefficients for each. 
These coefficients control the viscous torque and thus the mass accretion rate. 
We have not varied these coefficients or related physics, so it is not yet clear whether or how these would affect the truncation radius or the open flux for a given accretion rate (which is how our formulations are phrased). 

All of our simulations exist in the net spin-up regime. {Furthermore, MEs contribute solely spin-up torque to the star for all simulations. Therefore, we were unable to explore the possibility of MEs helping the star reach an equilibrium or net spin-down situation. Our torque formulations predict that increasing the magnetic field strength or decreasing the mass accretion rate (extrapolating beyond the parameter regime simulated) could push simulations into the propeller regime, where MEs could produce a spin-down contribution to the star and provide solutions that approach or even reach a net spin-down configuration. Alternatively (or in addition), more-massive stellar winds, i.e., higher stellar wind mass loss rates, may enable the transition into this net spin-down regime, and could be achieved by increasing the coronal temperature.}
There are also limitations to using a simple polytropic Parker wind to model our stellar winds, so their properties should be taken to be hypothetical. 
We are still unsure what mass loss rates are realistic for these stars, and what drives these stellar winds.
The outputted mass loss rates should not be taken as a true prediction of these simulations, but they are a result of unconstrained parameters, such as the coronal density, temperature, and adiabatic index. {Finally, there may be other mechanisms that are not captured by our simulations.}

{A proper modeling of the thermal structure of the disk should take into account a lot of effects (internal heating, stellar illumination and disk geometry, opacity, emissivity, dust content, dust-to-gas ratio; \citealp[see, e.g.,][]{doi:10.1146/annurev-astro-081309-130932}) that we neglected and replaced with a simple adiabatic assumption.
It is widely accepted that non-ideal MHD effects play a fundamental role in determining the accretion dynamics and the disk structure \citep[see, e.g.,][]{2014A&A...566A..56L,10.1051/0004-6361/201630056}, but we used a simpler alpha parameterization to control the gas-magnetic field coupling. }

The assumption of axisymmetry neglects 3D effects that may result in deviations of the open flux, the amount of flux participating in the MEs, and the size of the accretion flow, i.e., for a fixed accretion rate, there may be differences in the density within the funnel flow, the truncation radius, and thus the flow dynamics. 
To improve upon this, future studies must consider 3D SDI simulations. It is possible that the formulation developed in this paper can be applied to even multipolar cases if the dipolar component indeed determines the position of the truncation radius, i.e., if the dipole is stronger than any other multipole in the truncation region, and the multipoles do not introduce any closed loops at latitudes higher than the accretion spots, as the stellar flux available to accelerate a wind at latitudes higher than the accretion spots would be equivalent to that in our purely dipolar cases. However, to test this, future studies should consider more realistic magnetic field topologies, i.e., non-axisymmetry and additional multipolar components, and investigate the effectiveness of our torque formulations for these increasingly complex cases. 

\acknowledgments{LGI and SPM acknowledge funding from the European Research Council (ERC) under the European Union's Horizon 2020 research and innovation program (grant agreement No 682393; \textit{AWESoMeStars}: Accretion, Winds, and Evolution of Spins and Magnetism of Stars; \url{http://empslocal.ex.ac.uk/AWESoMeStars}). 

The authors would like to acknowledge the use of the University of Exeter High-Performance Computing (HPC) facility in carrying out this work. GP acknowledges funding from the European Research Council (ERC) under the European Union's Horizon 2020 research and innovation program (grant agreement No 742095; \textit{SPIDI}: Star-Planets-Inner Disk-Interactions); \url{https://www.spidi-eu.org}. 

We thank Andrea Mignone and others for the development and maintenance of the PLUTO code. Figures within this work are produced using the Python package Matplotlib \citep{2007CSE.....9...90H}.}

\software{Matplotlib \citep{2007CSE.....9...90H}, PLUTO \citep{0067-0049-170-1-228,2012ApJS..198....7M}.}

\bibliography{papers}

\appendix

\section{Parameterization of the mass accretion rate}\label{sec:param_Mdot}

Our formulation calculates the stellar torques for a specified or known mass accretion rate, so the formulations should be somewhat independent of the details of the physical processes and properties that determine $\dot M_\text{acc}$ in the disk. 
In our simulations, $\dot{M}_\text{acc}$ is a result of the particular initial disk profile, input parameters, and the physics included. 
In this appendix, we show how to predict $\dot{M}_\text{acc}$ from our simulation input parameters, although this may only apply for the initial disk conditions and physics that we have included in our simulations. The parameterization of $\dot{M}_\text{acc}$ allows for the truncation radius, and hence the accretion and stellar ME torque formulations, to be determined a priori from $\rho_\text{d}$. 

\subsection{Poloidal equilibrium derivation of the truncation radius}
The truncation radius can be derived from the poloidal equilibrium of the accretion disk and the stellar magnetic field. With $\rho$, $p$, and $v_R$ being the disk midplane density, thermal pressure and accretion speed, respectively, $c_\text{s} = (p/\rho)^{1/2}$ being the equatorial isothermal sound speed, and $m_s = |v_R/c_\text{s}|$ being the sonic Mach number of the accretion flow, this is expressed as

\begin{equation}\label{eq:trunc_rad_equil}
	\frac{B_\theta^2}{8\pi} = \frac{p + \rho v_R^2}{\beta_\text{tot}} = \frac{p(1+m_\text{s}^2)}{\beta_\text{tot}},
\end{equation}
where $B_\theta$ is the poloidal field at the disk midplane and the parameter $\beta_\text{tot}$ is of order unity. The usual plasma $\beta = 8 \pi p / B_\theta^2$ parameter is therefore

\begin{equation}\label{eq:beta}
	\beta = \frac{\beta_\text{tot}}{1+m_\text{s}^2}.
\end{equation}
Notice that here we are considering a region of the disk that is still accreting along the midplane. Our definition of the truncation radius corresponds to the position where the accretion flow starts to deviate and uplift to form the accretion columns, not to the radius at which the accretion flow is disrupted. In other words, our definition of the truncation radius corresponds to the intersection with the magnetic surface that envelops the accretion columns, not the one onto which the accretion columns lean on.

Taking the thermal height scale of the disk $H=Rc_\text{s}/v_\text{K} = c_\text{s}/\Omega_\text{K}$ where $v_\text{K} = (GM_\star/R)^{1/2}$ and $\Omega_\text{K}=v_\text{K}/R$ are the Keplerian toroidal and angular speeds, respectively, one can express the disk density and pressure as a function of the mass accretion rate $\dot{M}_\text{acc}$:

\begin{align}\label{eq:rho_Mdot}
\begin{gathered}	
\rho = \frac{\dot{M}_\text{acc}}{4\pi H R v_R}, \\
p = \rho c_\text{s}^2 = \frac{\dot{M}_\text{acc} v_\text{K}}{4\pi R^2 m_\text{s}^2}.
\end{gathered}
\end{align}
Therefore, Equation~(\ref{eq:trunc_rad_equil}) can be rewritten as

\begin{equation}\label{eq:trunc_rad_equil2}
	\frac{2 \dot{M}_\text{acc} v_\text{K}}{R^2 B_\theta^2} = \frac{\beta_\text{tot} m_\text{s}}{1+m_\text{s}^2},
\end{equation}
demonstrating that while the truncation can be seen as a poloidal pressure equilibrium, it also depends on the Mach number and accretion speed, which are determined by the angular momentum transport that drives accretion.

\subsection{Angular momentum conservation}

Using the $\phi$ component of the momentum equation (Equation~(\ref{eq:MHD})), we write the stationary angular momentum conservation equation, considering the torque due to a large-scale magnetic field and a viscous torque:

\begin{equation}\label{eq:ang_mom_cons_eq}
	\rho v_R \frac{d}{dR} \left(R^2 \Omega\right) = \frac{R B_\theta}{4\pi} \frac{d B_\phi}{dz} + \frac{1}{R} \frac{d}{dR} \left(R^2 \tau_{R\phi}\right),
\end{equation}
where $B_\theta$ and $B_\phi$ are the vertical and the toroidal components of the  magnetic field strength, respectively, and $\tau_{R\phi} = \rho \nu_\text{v} R (d \Omega / dr)$ is the $R\phi$ component of the viscous stress tensor (see Equation~(\ref{eq:stress})). By assuming an $\alpha$ parameterization \citep{1973A&A....24..337S} for the kinematic viscosity, $\nu_\text{v} = \alpha_\text{v} c_\text{s}^2 / \Omega_\text{K} = \alpha_\text{v} c_\text{s}^2 R / v_\text{K}$. In a Keplerian disk, $\tau_{R\phi} = - \alpha_\text{v} p$. We define the dimensionless quantity $Q_\text{rot} = v_\text{K}^{-1} d(R^2 \Omega)/dR$ (where $Q_\text{rot} = 1/2$ for a Keplerian disk). Therefore, Equation~(\ref{eq:ang_mom_cons_eq}) becomes

\begin{equation}\label{eq:ang_mom_cons_eq2}
	Q_\text{rot} \rho v_R v_\text{K} = \frac{R B_\theta}{4\pi} \frac{d B_\phi}{dz} - \frac{\alpha_\text{v}}{R} \frac{d}{dR} \left(R^2 p\right).
\end{equation}

Assuming $p = c_\text{s}^2 \rho$, $dB_\phi/dz \approx B_\phi^+/H$, $q=-B_\phi^+/B_\theta$, $(\alpha_\text{v}/R) d(R^2 p)/dR \approx \alpha_\text{v}$, and $\epsilon=H/R=c_\text{s}/v_\text{K}$, and $m_\text{s} = \lvert v_R \rvert /c_\text{s}$, we can rewrite Equation~(\ref{eq:ang_mom_cons_eq}) as

\begin{equation}\label{eq:ang_mom_cons_eq3}
	m_\text{s} = \left\lvert \frac{2 q}{Q_\text{rot} \beta} + \alpha_\text{v} \epsilon \right\rvert,
\end{equation}
where $\beta = 8 \pi p / B_\theta^2$. While viscous accretion is usually strongly subsonic (typically $\alpha \epsilon \ll 1$), an equipartition field, with both $q$ and $\beta$ of order unity, tends to yield a sonic accretion velocity. 
Since these are the typical conditions in the truncation region, it is safe to neglect the viscous torque:

\begin{equation}\label{eq:ang_mom_cons_eq4}
	m_\text{s} \simeq \frac{2 \lvert q \rvert}{Q_\text{rot} \beta}.
\end{equation}
.
\subsection{Mass accretion rate scaling}

If we assume the typical SDI parameterization for $q$:

\begin{equation}\label{eq:q_appendix}
	q = K_q \left(\frac{\Omega_\star}{\Omega} - 1\right),
\end{equation}
where $K_q$ is a best-fit dimensionless parameter, and neglect the viscous torque, we can combine Equations~(\ref{eq:beta}),~(\ref{eq:trunc_rad_equil2}), and~(\ref{eq:ang_mom_cons_eq4}) and get the following set of equations:

\begin{align}\label{eq:q_appendix2}
	\begin{gathered}
      \frac{2 \dot{M}_\text{acc} v_\text{K}}{R^2 B_\theta^2} = \frac{\beta_\text{tot} m_\text{s}}{1+m_\text{s}^2}, \\
      \frac{K_q}{4 \pi Q_\text{rot}} \frac{B_\theta^2}{p} \left\lvert\frac{\Omega_\star}{\Omega} - 1\right\rvert = m_\text{s}.      
    \end{gathered} 
\end{align}
$\beta_\text{tot}$ is plausibly of order unity, since the magnetic field should be able to stop the disk accretion flow. Sonic accretion, together with a magnetic field that can balance both the ram and thermal disk pressure, seems to represent $m_\text{s}$ of order unity as an optimal condition to truncate the disk. Sonic accretion, along with an equipartition poloidal field, also corresponds to a field twist $q \lesssim 1$, which is compatible with the maximum twist allowed to avoid the inflation of the closed magnetosphere. We notice also that the factor $m_\text{s}/(1+m_\text{s}^2)$ has a maximum at $m_\text{s}$, meaning a $m_\text{s}=1$ condition roughly corresponds to the outermost position where the conditions for truncation are met.

\begin{figure}
\begin{center}
\includegraphics[width=0.4625\textwidth]{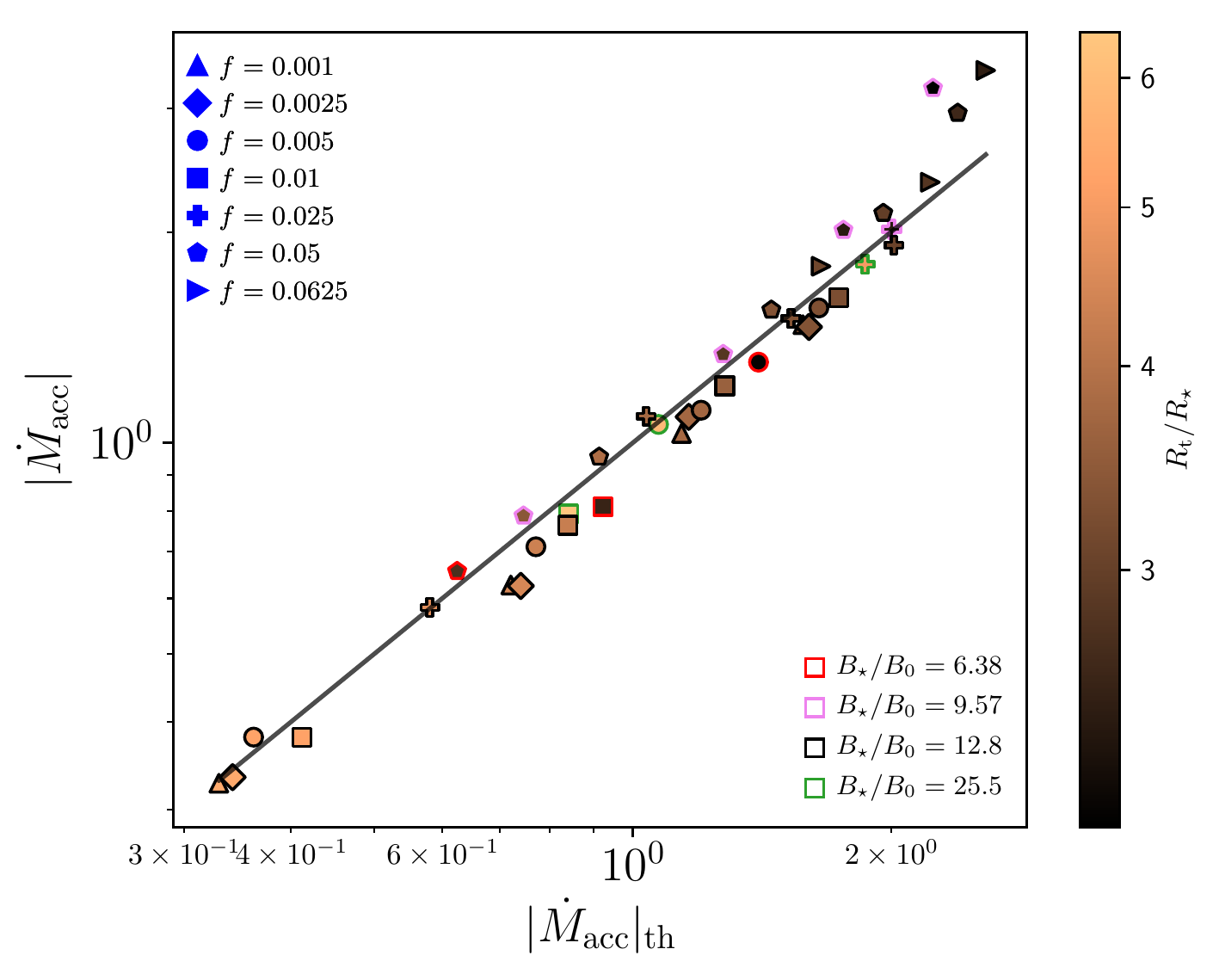}
\caption{Outputted $\lvert\dot{M}_\text{acc}\rvert$ as a function of the parameterization in Equation~(\ref{eq:Mdot_acc_param}) for all simulations. The black line shows $y=x$, indicating goodness of fit, giving $K_M=0.00221$.\label{fig:Mdot_acc_v_Mdot_acc_theory}}
\end{center}
\end{figure}

The first expression in Equation~(\ref{eq:q_appendix2}) provides the familiar relation for the truncation radius $R_\text{t}=R_\text{t}(\dot{M}_\text{acc},B_\star)$ (see Equation~(\ref{eq:Rt_Y_acc})), whilst the second provides an implicit relation for the truncation radius $R_\text{t}=R_\text{t}(Q_\text{rot}(R_\text{t}),\Omega_\star,\Omega(R_\text{t}))$. At the truncation radius, we assume that the thermal disk pressure scales with the initial disk pressure profile (Equation~(\ref{eq:rho_disk_initial})), i.e., $p(R_\text{t}) \propto \rho_{\text{d,}\star} v_\text{esc}^2 (R_\text{t}/R_\star)^{-5/2}$ (for fixed disk aspect ratio $\epsilon$), $B_\theta(R_\text{t}) \simeq B_\theta(R_\text{t}) = K_\text{B} B_\star (R_\text{t}/R_\star)^{m_\text{B}}$ (Equation~(\ref{eq:B_theta_Rt})), and set $\Omega_\star/\Omega(R_\text{t}) = (R_\text{t}/R_\text{co,m})^{3/2} = (f/K_\text{acc}) (R_\text{t}/R_\star)^{(3/2)-m_\text{acc}}$ (Equation~(\ref{eq:Rco_eff})), giving

\begin{align}\label{eq:Rt_appendix}
     \frac{R_\text{t}}{R_\star} = \frac{m_\text{s}' Q_\text{rot}}{2 K_q K_\text{B}^2} \left\{\left\lvert\frac{f}{K_\text{acc}} \left(\frac{R_\text{t}}{R_\star}\right)^{(3/2)-m_\text{acc}} - 1\right\rvert \left(\frac{B_\star^2}{4 \pi \rho_{\text{d,}\star} v_\text{esc}^2}\right)\right\}^{-2 / (5+4m_\text{B})}.     
\end{align}
The $m_\text{s}'$ factor does not correspond to the actual Mach number in the evolved solution, due to the assumptions made above. 
Using our $R_\text{t}$ parameterization (Equations~(\ref{eq:Rt_Y_acc})-(\ref{eq:Y_acc})), a parameterization of the mass accretion rate for our simulations can be written as

\begin{align}\label{eq:Mdot_acc_param}
     \lvert \dot{M}_\text{acc} \rvert = K_M \frac{B_\star^2 R_\star^2}{4\pi v_\text{esc}} \left\{\left\lvert\frac{K_\text{t}^{(3/2)-m_\text{acc}} f}{K_\text{acc}} \left(\frac{B_\star^2 R_\star^2}{4 \pi \lvert \dot{M}_\text{acc} \rvert v_\text{esc}}\right)^{m_\text{t}[(3/2)-m_\text{acc}]} - 1\right\rvert^{-1} \frac{4 \pi \rho_{\text{d,}\star} v_\text{esc}^2}{B_\star^2}\right\}^{-2 / [m_\text{t} (5+4m_\text{B})]}, 
\end{align}
where $K_M = [m_\text{s}' Q_\text{rot} / (2 K_q K_\text{B}^2 K_\text{t})]^{-1/m_\text{t}}$ is a best-fit dimensionless parameter. This is a transcendental equation that requires to be solved iteratively. In Figure~\ref{fig:Mdot_acc_v_Mdot_acc_theory}, we plot the outputted mass accretion rate as a function of the theoretical mass accretion rate calculated using Equation~(\ref{eq:Mdot_acc_param}), and show excellent correspondence between these with best-fit dimensionless parameter $K_M = 0.00221$.

\section{Using our formulation to calculate net torque}\label{sec:stellar_evo_code}

For clarity, here is a simple guide to using the torque formulations derived in this study.

\begin{enumerate}
	\item{Specify the following input parameters: stellar mass ($M_\star$), stellar radius ($R_\star$), surface polar magnetic field strength ($B_\star$), stellar break-up fraction ($f$), mass accretion rate ($\dot{M}_\text{acc}$), and stellar wind mass loss rate ($\dot{M}_\text{wind}$). Alternatively, one can use Appendix~\ref{sec:param_Mdot} to specify the initial stellar surface disk density ($\rho_{\text{d,}\star}$) rather than the mass accretion rate.}
	\item{Calculate the truncation radius ($R_\text{t}$), using Equations~(\ref{eq:Rt_Y_acc})-(\ref{eq:Y_acc}).}
	\item{Calculate the accretion torque ($\dot{J}_\text{acc}$), stellar ME torque ($\dot{J}_{\text{ME},\star}$), and stellar wind torque ($\dot{J}_\text{wind}$) using Equations~(\ref{eq:acc_torque_2}),~(\ref{eq:ME_torque_2_Rt}), and~(\ref{eq:full_sw_torque}), respectively.}
	\item{Sum each torque contribution to calculate the net stellar torque ($\dot{J}_\star = \dot{J}_\text{acc} + \dot{J}_{\text{ME},\star} + \dot{J}_\text{wind}$).}
\end{enumerate}

SDI simulations in this study, and in general, are performed over several dynamical timescales, but evolutionary timescales of these stars are on the order of millions of years. 
Therefore, to predict the spin evolution of these stars, one can determine the net stellar torque for a star with given global properties, using torque formulations derived from these SDI simulations, and then evolve this over time using a stellar evolution code. 
However, it is important to consider the validity of our torque formulation in regimes not considered by the parameter study; for example, if the truncation radius becomes smaller than the stellar radius, i.e., $R_\text{t} < R_\star$ (outside the magnetic accretion regime), we recommend enforcing a limit of $R_\text{t} \ge R_\star$.
On the other hand, if $R_\text{t}$ exceeds the magnetic corotation radius ($R_\text{t} > R_\text{co,m}$, i.e., the net spin-up stellar torque regime), our physically-motivated formulations can be extrapolated and used here, but doing so will be uncertain because real systems may transition to different regimes in parameter ranges far outside what we have explored.  The situation is even more unclear for the propeller regime (where $R_\text{t}>R_\text{co}$), where the accretion flow is expected to be significantly disrupted.

\section{spin-equilibrium solution for our formulation}\label{sec:spin_eq}

By setting the sum of the accretion (Equation~(\ref{eq:acc_torque_2})), stellar ME (Equation~(\ref{eq:ME_torque_2_Rt})), and stellar wind (Equation~(\ref{eq:full_sw_torque})) torques equal to zero, i.e., $\dot{J}_\text{wind} + \dot{J}_\text{acc} + \dot{J}_{\text{ME,}\star} = 0$, one can express the following transcendental equation (using best-fit dimensionless parameters in Table~\ref{tab:dimensionless_params}):

\begin{align}\label{eq:spin_eq_sol}
\begin{split}
58.2 \left(\frac{\dot{M}_\text{wind}}{\lvert \dot{M}_\text{acc} \rvert}\right)^{0.254} f_\text{eq}^{1.09} \left[1 + \left(\frac{f_\text{eq}}{0.0356}\right)^2\right]^{-0.373} \left(\frac{R_\text{t}}{R_\star}\right)^{0.191} - 0.775 \left(\frac{R_\text{t}}{R_\star}\right)^{0.353} + 0.312 \left[\frac{f_\text{eq}}{0.775} \left(\frac{R_\text{t}}{R_\star}\right)^{1.65} - 1\right] \left(\frac{R_\text{t}}{R_\star}\right)^{0.402} = 0,
\end{split}
\end{align}
where $f_\text{eq}$ is the spin rate (as a fraction of break-up) for the star to have a net zero torque, i.e., the ``equilibrium spin rate.''  
To calculate $f_\text{eq}$, for a given $R_\text{t}$ and $\dot{M}_\text{wind}/\lvert\dot{M}_\text{acc}\rvert$, this expression must be solved iteratively. 
Alternatively, if the spin rate is specified/fixed, one can instead solve for the predicted equilibrium mass ejection efficiency for a given $R_\text{t}$ (or vice versa).

\end{document}